\documentclass[%
 aip,
 amsmath,amssymb,
 reprint,%
]{revtex4-1}

\usepackage{graphicx}% Include figure files
\usepackage{dcolumn}% Align table columns on decimal point
\usepackage{bm}% bold math
\usepackage{amssymb, amsmath, amsthm}
\usepackage[utf8]{inputenc}
\usepackage[T1]{fontenc}
\usepackage{mathptmx}
\usepackage{etoolbox}
\usepackage{siunitx}
\usepackage{xcolor}

\usepackage{subcaption}
\usepackage{tikz}

\usetikzlibrary{calc,trees,positioning,arrows,chains,shapes.geometric,%
    decorations.pathreplacing,decorations.pathmorphing,shapes,%
    matrix,shapes.symbols}
    
\usetikzlibrary{shapes.arrows}

\usepackage{forest}

\definecolor{folderbg}{RGB}{124,166,198}
\definecolor{folderborder}{RGB}{110,144,169}

\def\Size{4pt}
\tikzset{
  folder/.pic={
    \filldraw[draw=folderborder,top color=folderbg!50,bottom color=folderbg]
      (-1.05*\Size,0.2\Size+5pt) rectangle ++(.75*\Size,-0.2\Size-5pt);  
    \filldraw[draw=folderborder,top color=folderbg!50,bottom color=folderbg]
      (-1.15*\Size,-\Size) rectangle (1.15*\Size,\Size);
  }
}

\newcommand{\pa}{\partial}

\makeatletter
\def\@email#1#2{%
 \endgroup
 \patchcmd{\titleblock@produce}
  {\frontmatter@RRAPformat}
  {\frontmatter@RRAPformat{\produce@RRAP{*#1\href{mailto:#2}{#2}}}\frontmatter@RRAPformat}
  {}{}
}%
\makeatother
\begin{document}

\preprint{AIP/123-QED}

\title[]{The destabilising effect of particle concentration in inclined settlers}
% Force line breaks with \\
\author{Cristian Reyes}
\affiliation{ 
Laboratory for Rheology and Fluid Dynamics, Department of Mining Engineering, University of Chile. Beauchef 850, 8370448 Santiago, Chile
}%
 \email{cristian.reyes@ug.uchile.cl}
\author{Cristobal Arratia}%
\affiliation{%
Nordita, Stockholm University and KTH Royal Institute of Technology, \\ SE-106 91 Stockholm, Sweden
}%

\author{Christian F. Ihle}
% \homepage{http://www.Second.institution.edu/~Charlie.Author.}
 \affiliation{ 
Laboratory for Rheology and Fluid Dynamics, Department of Mining Engineering, University of Chile. Beauchef 850, 8370448 Santiago, Chile
}%

\date{\today}% It is always \today, today,

\begin{abstract}
Water scarcity has required constant water recycling, leading to a decline in water quality, further exacerbated by high concentrations of fine particles that reduce the efficiency of solid-liquid separation systems. Inclined settlers offer a viable secondary treatment option for high-turbidity water. Effective design requires understanding of operational conditions, geometry, and suspension properties. Using OpenFOAM, computational fluid dynamics simulations were performed for a continuous inclined countercurrent conduit to assess the influence of inlet particle concentration on efficiency, exploring various Surface Overflow Rates (SOR) and inclination angles. The results show that the steady state in which the flow settles is strongly dependent on the particle concentration.  For very low particle concentrations, the flow is mostly stationary with little to no resuspension of particles. Increasingly unstable regimes are observed to emerge as the inlet concentration increases, leading to increased particle resuspension. 
   Instabilities arise from overhanging zones at the tip of the suspension, generating recirculation zones that enlarge the resuspension region and induce entrainment within the bulk suspension. Shear instabilities become noticeable at large particle concentrations, further increasing resuspension.
   Different regimes were identified, influenced by the SOR and the inclination angles. 
   Additionally, a Reynolds number characterizing these systems is proposed alongside a scale analysis. The findings highlight particle concentration as a critical parameter in inclined plate settler design.
\end{abstract}

\maketitle

\section{Introduction}

Solid-liquid separation is a crucial process in water recovery, with its efficiency highly dependent on the type and concentration of particles involved. For example, in the mining industry, where the water footprint has become a central issue, the emergence of clays within gangue minerals has become a significant impediment to maintaining the quality of recycled water \citep{connelly2011high, grafe2017clays}. Inclined settler technology presents itself as a viable and environmentally friendly option for low footprint gravitational secondary physical treatment of turbid water. These systems serve the purpose of clarification either as standalone processes or in conjunction with others, depending on the nature of the suspensions undergoing treatment. 
Falling under the classification of high-rate settlers, inclined settler technology is predominantly utilized for water clarification, often in tandem with pretreatment operations such as coagulation and flocculation. This combined approach is a cost-effective solution for secondary treatment in wastewater plants \citep{clark2009inclined, edzwald2011water}. The operating mode of these systems is mainly continuous, as observed in the water treatment industry \citep{edzwald2011water}. In continuous operation, inclined settlers are typically categorized as countercurrent or cocurrent and occasionally as cross-flow, based on the relative positioning of the feed and the underflow. In particular, countercurrent equipment, where the feed is located at or near the bottom, represents the most prevalent configuration in inclined settler equipment \citep{wilson2005introduction, crittenden2012mwh}.
A fundamental aspect of this technology is the use of inclined confining elements. Among the most common types are Lamellar settlers, a class of physical separators that employ an array of inclined plates to facilitate the reorganization between the solid and liquid phases, leveraging the Boycott effect \citep{boycott1920sedimentation}. Lamella settlers, as inclined element settlers in general, are recognized for their high efficiency and relatively low footprint in particle separation, especially at low concentrations \citep{tarpagkou2014influence}. They find widespread application in wastewater contexts \citep{salem2011approach}. \citet{reyes2022review} provides a recent review on the subject of inclined settlers.

 Concerning design, the prevalent method used to determine the size of inclined settler components assumes that particles follow either a linear or parabolic path from any position perpendicular to the bottom wall of the inclined cell, depending on the type of flow and the geometry \citep{Yao70JotWPCF, letterman1999water}. This assumption relies on the system being within the dilute limit, maintaining steady-state conditions in each inclined flow conduit, and the flow being uniform and reasonably represented by the mean flow velocity. Despite the arguments of \citet{Fadel90JoEE} that the constant velocity model underestimates tube length, this approach continues to be widely endorsed in the literature (e.g., studies of \citet{letterman1999water, wilson2005introduction, crittenden2012mwh, Lin14book} and \citet{droste2018theory}). 
 Perhaps surprisingly, the criteria thus derived for settler size design coincide with a different size determination based on a mass balance approach known as PNK \citep{ponder1925sedimentation,nakamura1937cause}, later theoretically justified and extended to continuous operation by \citet{acrivos1979enhanced}. According to these modeling approaches, the suspended particles can reach a position $L_m$ (given below in \S \ref{sec:SystemDescription}) along the settler.

However, these designs overlook the impact of hydrodynamic instabilities on the efficiency of the final system design. Instability occurs within the cell interior when flow inertia exceeds viscous forces, forming two- or three-dimensional flow structures. These structures can impede settling by lifting particles and potentially carrying them toward the settler's overflow. This mechanism significantly contributes to the degradation of the total removal efficiency (TRE). Instabilities in these inclined systems have been studied mathematically, using linear stability analysis \citep{herbolzheimer1983stability, davis1983wave, shaqfeh1986effects, borhan1988sedimentation}, and numerically, using finite difference, finite volume, CFD-DEM or lattice Boltzmann methods \citep{laux1997computer, patankar2001modeling, chang2019three, zhang2021fluid, esipov20232d}. 
These analyses have focused on the development of shear instabilities around the interface between the clear water layer next to the upper wall and the suspension. 
\citet{laux1997computer} noted that with increasing particle concentration, their findings diverged from the predictions outlined in the PNK model (\citet{borhan1988sedimentation} also observed this effect; see also \citep{reyes2022review}, for a review). Designing these inclined settlers correctly and understanding the physics inside is essential to optimize these processes.

In the present paper, we report on two-dimensional numerical simulations of the flow in an inclined container that mimics a single-cell settler. Although such numerical simulations are not an exact representation of the three-dimensional effects that occur at both the inlet and the discharge of industrial equipment, they give the essential features of the underlying physical mechanism that degrades the TRE in them. Starting from very dilute suspensions, we studied the effect of particle concentration on resuspension and efficiency within the inclined settler, observing a marked degradation of performance as concentration increases. 
We provide a detailed description of different regimes that we identify along the destabilization route through which efficiency breaks down.
We characterize this destabilization route for different angles and feed rates. Although we consider monodisperse suspensions, we end with an exploration of different particle sizes.

\section{System description}
\label{sec:SystemDescription}

\begin{figure}[!ht]
\centering
\includegraphics[width=10cm]{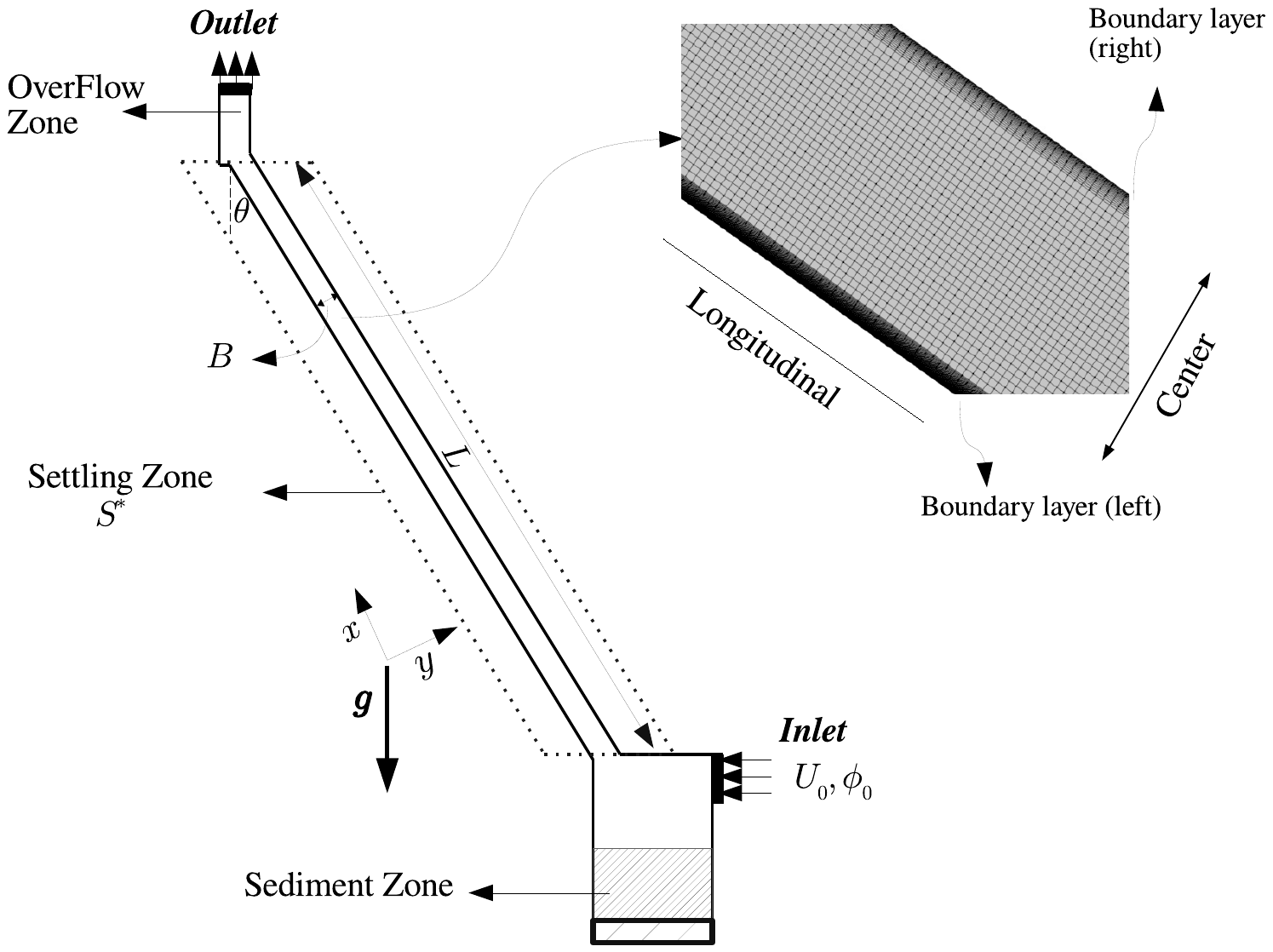}
\caption{Geometry used in numerical simulations. The main geometrical and flow parameters are shown. In the close-up area the structure of the mesh used can be appreciated.}\label{f:system}
\end{figure}

Figure~\ref{f:system} illustrates the geometry under examination along with the mesh structure within the settling zone. The setup represents a countercurrent inclined settler featuring plates characterized by a length $L=1$ m, plate separation $B=2$ cm, and an inclination angle $\theta$ from the vertical.  This inclined cell is connected to an inlet and a hopper in its lower part and an outlet zone in its upper part.  
In this system a suspension with particle size $d$ and volume concentration $\phi_0$ is injected with a velocity $U_{0}$ to settle the particles between the plates.
The inlet size $l_{in}=B/2$ is half the width of the cell and is slightly separated from the beginning of the cell to avoid any effect of the inflow condition in the cell. Similarly, the outlet is separated from the cell to avoid backflow. The excess of accumulated particles is removed from the hopper zone as described in Section \ref{sec:BCs}, below. The coordinates $x$ and $y$ correspond to the longitudinal and transverse directions to the walls of the conduit.

As mentioned above, the models used to design these inclined settling tanks \citep{Yao70JotWPCF,letterman1999water, ihle2022new} turn out to coincide \citep{reyes2022review} with the continuous extension of the PNK theory \citep{acrivos1979enhanced} which is valid under certain suspension and flow criteria \citep{davis1985sedimentation}. According to these models, the particles in the suspension can raise along the settler until reaching an interface position given by
\begin{equation} \label{eq:LmDef}
    L_m = \left(\frac{B}{\sin\theta}\right)\left(\frac{\bar{U}}{W_s} - \cos\theta\right),
\end{equation}

where $\bar{U}=l_{in}U_{0}/B$ is the nominal mean velocity in the cell and $W_s$ is the Stokes settling velocity. 
In other words, $L_m$ is the theoretical longitudinal length of the suspension layer (measured from the bottom) inside the cell.

\subsection{Input parameters}

\subsubsection{Surface overflow rate (SOR)}
A common design parameter that describes the capacity of the settler is the surface overflow rate (SOR), which relates the flow in the settler (operational parameter) to the projected area (geometrical parameter). For an inclined settler this parameter is calculated as \citep{clark2009inclined}
\begin{align} \label{eq:SORDef}
    \text{SOR} = \frac{\bar{U}}{\frac{L}{B}\sin\theta + \cos\theta}.
\end{align}

The SOR corresponds to the minimum settling velocity that this settler can theoretically remove (as can be seen by replacing the settler length $L$ in \eqref{eq:SORDef} by the predicted interface position $L_m$ in \eqref{eq:LmDef}). Thus, SOR should satisfy $\text{SOR} < W_s$, in order to prevent the conduit from completely filling with particles.

Consequently, when studying different particle sizes, an important parameter to guide comparisons is the ratio of SOR to settling velocity $\frac{\text{SOR}}{W_s}$. This ratio is directly related to the fraction of the length cell $L$ that would be occupied by the suspension (in the ideal case). In fact, when the ratio $L/B$ is large, it is satisfied that:

\begin{equation}\label{eq:SORbyWS}
    \frac{\text{SOR}}{W_s} \approx \frac{L_m}{L}.
\end{equation}

\subsubsection{Physical parameters of the simulations}

We study the effect of particle concentration considering angles used in inclined settlers and particle sizes found in thickener supernatants in mining operations. 
In this study, the simulations were performed in two dimensions, which implies that three-dimensional structures that could potentially develop are not modeled. Nonetheless, as \citet{chang2019three} noted, even in a system with larger particles (100~\si{\micro\meter}) and a non-dilute concentration ($\phi_0=\SI{10}{\%}$), the perturbations behaved in a quasi-2D manner. Additionally, \citet{laux1997computer} obtained accurate results in 2D simulations that exhibited instabilities similar to those observed in experiments.

The different values of particle concentration $\phi_0$, SOR and angle of inclination ($\theta$) for our main study with particle diameter $d=\SI{10}{\micro m}$ are given in Table \ref{t:mainParams}. Particle concentration, in this study, represents the volumetric fraction of particles within the system. In our two-phase simulation model, it is calculated as the ratio of particle volume to total volume within a control volume (here, the control volume corresponds to a computational cell). The evolution of particle concentration over time and space is then determined using equation~\eqref{eq:part_calc}.
We perform simulations for all possible combinations of these parameters and put an emphasis on the effect of particle concentration. 

\begin{table}[!ht]
\centering
\caption{Range of values used in the 90 numerical simulations with particle diameter $d=\SI{10}{\micro m}$ constituting the main part of our study.}
\label{t:mainParams}
\begin{ruledtabular}
      \begin{tabular}{ p{3cm}p{5cm}  }
     Parameter & Values\\
     \hline
     Inlet particle concentration $\phi_0$ (\SI{1e-3}{}) & \SI{0.02}{}, \SI{0.05}{}, \SI{0.1}{}, \SI{0.125}{}, \SI{0.15}{}, \SI{0.2}{}, \SI{0.4}{}, \SI{0.8}{}, \SI{1.2}{}, \SI{2}{}   \\
     SOR (mm/s) & 0.018, 0.036, 0.054 \\
     Inclination angle $\theta$ & 35, 45, 55° \\
    \end{tabular}  
\end{ruledtabular}
\end{table}

To get an idea of the expected behavior of the operating settler according to \eqref{eq:SORbyWS}, the SOR values can be compared with the Stokes sedimentation velocity $W_s$. The value of $W_s$ for the main case, as well as the values of $W_s$ for the exploratory cases with particle diameters $d=\SI{5}{\micro m}$ and $d=\SI{20}{\micro m}$ reported in \S \ref{sec:diameterEffect}, are given in Table \ref{t:diffdParams}. 
Table \ref{t:diffdParams} also gives the range of nominal Reynolds numbers $Re=\bar{U} B/\nu$ (with $\nu$ the kinematic viscosity of water) of the main study, as well as the values of $Re$ for exploratory cases with different $d$. 

Another important nondimensional parameter is $\Lambda$ defined by \citet{acrivos1979enhanced} as the ratio between the sedimentation Grashof number and the sedimentation Reynolds number $Re_s=W_s\mathcal{L}/\nu$, which gives
\begin{equation}\label{eq:lambda}
\Lambda = \frac{gR_{\rho} \mathcal{L}^3/\nu^2}{Re_s}=\frac{9}{2}\left(\frac{\mathcal{L}}{a}\right)^2\phi    
\end{equation}
where $R_{\rho}=\phi_0(\rho_s-\rho_f)/\rho_f$ is the density ratio between the suspension and the clear fluid, $\mathcal{L}$ is a length scale of the flow (with $L_m$ used in this study),  and $a=d/2$ is the particle radius. \citet{acrivos1979enhanced} show that the PNK theory consistent with \eqref{eq:LmDef} is recovered in the limit as $\Lambda\rightarrow\infty$. 
Different asymptotic similarity solutions have been obtained by \citet{shaqfeh1986effects} depending on the value of $\mathcal{R}^{1/2}$ as the limit $\Lambda\rightarrow\infty$ is approached, where $\mathcal{R}=Re_s\Lambda^{-1/3}$. This family of solutions bridges the gap between a viscous regime for $\mathcal{R}<<1$ and an inertial regime for $\mathcal{R}>>1$. 
Which of these similarity solutions, if any, is approached by the actual (experimental or numerical) solution to the full problem is not known {\it a priori}. 
The values of $\Lambda$ for our simulations are shown in Table \ref{t:diffdParams}. In practical cases, this parameter has large values ($>10^6$, \cite{Blanchette03thesis}). 

Inlet velocities for particles with diameters of 5 and \SI{20}{\micro m} were calculated by maintaining a constant ratio $\frac{SOR_{10
}}{W_{s, 10}} = \frac{SOR_{5}}{W_{s,5}} = \frac{SOR_{20}}{W_{s,20}}$ , with subscripts denoting particle diameter in micrometers.
This approach was adopted because application of the same SOR would result in excessively high inlet velocities for the \SI{5}{\micro m} particles and insufficiently low inlet velocities for the \SI{20}{\micro m} particles, potentially leading to non-representative characterizations of the settling dynamics.

\begin{table}

\centering
\caption{Range of values used in the numerical simulations. \emph{main} means the values used for the simulations associated with 10 micron particles.}
\label{t:diffdParams}

\begin{ruledtabular}
    \begin{tabular}{ p{2cm}p{1.8cm}p{3cm}p{2cm} }
     Parameter & $d=\SI{5}{\micro m}$ &  $d=\SI{10}{\micro m}$ (\emph{main}) &   $d=\SI{20}{\micro m}$\\
     \hline
     $W_s$ (mm/s) & 0.023 & 0.093 & 0.371   \\
     Re ($\bar{U} B/\nu$)& 6.5 & 10.7-45 & 198 \\
     $\Lambda$   & \SI{1.44e9}{} & \SI{3.6e6}{} - \SI{3.6e8}{} & \SI{9e5}{} \\
     %Distance between plates & 2 cm\\
    \end{tabular}
    \end{ruledtabular}
\end{table}

Another relevant feature of the behavior of the studied system is related to the fact that the profile is independent of the length of the inclined cell. We tested the same case for different geometries where only the cell length was modified. According to the results of the simulations, changing the length does not generate relevant changes in the concentration profile for x < L, for the same values of inlet velocity $U_0$ and particle concentration $\phi_0$. Thus, the results obtained would be valid for any cell smaller than L (equal to 1 m in this article), as long as the same boundary conditions are considered.

\subsection{Performance parameters}
\subsubsection{Local turbidity removal efficiency}

The turbidity removal efficiency is a parameter widely used to measure the overall degree of solid-liquid separation achieved by the cell. A local value of the separation efficiency at a specified height x (measured from the base) is defined as:
\begin{align}
    \text{TRE}_{local} = \frac{\phi_0 - \bar{\phi} (x)}{\phi_0}
\end{align}

where $\phi_0$ is the concentration of particles in the inlet and $\bar{\phi} (x)$ is the average concentration in a perpendicular coordinate segment within the cell (in the normal direction of the wall). Measuring this parameter locally allows us to analyze the length where the desired efficiency is obtained given a flow rate, and to relate the TRE with a measure of effective length to achieve a particular particle concentration at the overflow.  Despite the importance of this parameter, the potential objective of completely eliminating particles from the overflow would generally require extremely long settling elements. This trade-off between process objective fulfillment and cost requires tolerating some degree of turbidity at the overflow.

\subsubsection{Interface Position}

In continuous operation, the interface position is the location along the inclined settler at which the average particle concentration reduces below a prescribed value. For these simulations, two different positions were considered. The first corresponds to the position of the bulk suspension, $L_b$, and the second corresponds to the maximum position of the resuspension zone, $L_r$. For the bulk suspension interface, we calculate the position where $\bar{\phi} (x)$ is closest to the inlet particle concentration value ($\phi_0$).
On the other hand, the resuspension interface $L_r$ was calculated as the position where the mean concentration corresponds to $1\%$  of $\phi_0$.

\subsubsection{Mass overflow parameter} \label{sec:massOverflow}

This parameter allows us to compare the behavior of the system with the theoretical prediction of the interface position $L_m$ given by \eqref{eq:LmDef} commonly used to design these inclined settling tanks. 
Thus, the overflow of solid mass beyond the theoretical prediction, `mass overflow parameter' ($mo$) for short, % and mass under ($mu$) this height, 

is defined as
\begin{align} 
    mo =
    \int_{L_m}^L \rho_s\phi dS^*,
\end{align}
where $\rho_s$ is the solid density. As the problem at hand is effectively two-dimensional, the integration is over a surface $S^*$ above $L_m$ within the settling zone. Therefore, $mo$ has units of mass per unit length. %The full mass in this zone would be calculated with the same integral from 0 to $L$. 

\subsubsection{Mean thickness of the clear water layer}

In this type of system, three layers associated with different particle concentrations develop: (i) the clear water layer with $\phi = 0$, (ii) the suspension layer with a concentration roughly equal to the inlet concentration, and (iii) the sediment layer with a high concentration that forms a slowly flowing layer at the bottom of the cell.  In countercurrent systems below the bulk interface, the clear water layer is smaller than the suspension layer, while for low-concentration cases, the sediment layer is submillimetric and can be neglected. Therefore, the analysis focuses only on the suspension and the clear water layers. 
As the thickness of the suspension is approximately $\delta_{susp}(x) \approx B - \delta_{cw}(x) $, only the behavior of the thickness of the clear water layer, $\delta_{cw}(x)$, will be reported. This parameter allows us to characterize how the particle concentration profile develops in the direction perpendicular to the plates.  Furthermore, according to the theory proposed by \citet{leung1983lamella}, this thickness is sensitive to the concentration of feed particles and has a direct relationship with the balance of forces within the fluid.
The mean thickness of the clear water layer was calculated by the following integral:

\begin{align}
    \bar{\delta} = \frac{1}{L_b}\int_0^{L_b} \delta_{cw}(x) dx,
\end{align}
where $L_b$ is the bulk suspension position. The local thickness $\delta_{cw}$ refers to the perpendicular distance from the upper wall to the sediment layer. It is calculated by identifying the position where the concentration of particles falls below a predefined threshold value, specifically $0.4\phi_0$. This approach is adopted due to the diffuse nature of the layer; employing a threshold value equal to $\phi_0$ would inaccurately delineate the position of the interface.

\section{Governing equations and numerical implementation}

The performance parameters referred to in the previous section are obtained from high resolution two-dimensional numerical simulations of a two phase model, following the Euler-Euler approach (i.e., both the solid and liquid phases are assumed to be continuous), using the OpenFOAM library \citep{reyes2016stability, reyes2019heat}. OpenFOAM solves governing equations using the Finite Volume Method. In this approach, the computational domain is divided into small control volumes, and the integral form of conservation equations is applied to each, ensuring conservation laws are met.

The Eulerian model was selected over a Lagrangian approach due to computational constraints. While particle concentrations are low, the small particle size (\SI{10}{\micro m}) results in a high particle count, following \(N_p \sim 1/V_p\) and \(V_p \propto d^3\). This makes a Lagrangian simulation excessively expensive, hence the Eulerian choice.

In this two phase model the corresponding momentum equations are as follows:
\begin{equation}\label{Eq:U1}
\begin{aligned}
            \frac{\partial (\phi_1\rho_1 \textbf{U}_1)}{\partial t} + \nabla\cdot \left(\phi_1\rho_1 \textbf{U}_1\textbf{U}_1\right) &= -\phi_1\nabla p_{rgh} + \phi_1\rho_1 \textbf{g} \\ &+\nabla\cdot\left(\phi_1\bm{\tau}_1\right) - p_s\nabla\phi_1 \\ &+ \textbf{F}_1, 
\end{aligned}
\end{equation}

\begin{equation}\label{Eq:U2}
\begin{aligned}
    \frac{\partial (\phi_2\rho_2 \textbf{U}_2)}{\partial t} + \nabla\cdot \left(\phi_2\rho_2 \textbf{U}_2\textbf{U}_2\right) &= -\phi_2\nabla p_{rgh} + \phi_2\rho_2 \textbf{g} \\
    &+ \nabla\cdot\left(\phi_2\bm{\tau}_2\right) + \textbf{F}_2, 
\end{aligned}
\end{equation}

where $\phi_0$, $\rho_i$ and $\textbf{U}_i$  are the volume fraction, density, and velocity of each phase. Here, 1 and 2 represent the solid and water phases, respectively, where it must be satisfied that $\phi_2 = 1 - \phi_1$. $\textbf{F}_1 = -\textbf{F}_2$ are inter-phases forces, $p_s$  is a solid pressure that models interparticle contact forces, and $p_{rgh}$ is a common pressure that ensures incompressibility. 
\begin{align}
    \nabla\cdot \textbf{V}= 0,
\end{align}
where
\begin{equation}\label{Eq:inc}
\textbf{V}\equiv\phi_1\textbf{U}_1 + \phi_2\textbf{U}_2,
\end{equation}
is the (volumetric) mean velocity. Equations~\eqref{Eq:U1}, \eqref{Eq:U2} and \eqref{Eq:inc} are used to derive a pressure-velocity coupled equation, which is employed to correct the velocities of both phases. The two-phase model employs the Rhie-Chow method \citep{rhie1983numerical} to resolve pressure-velocity coupling. This standard projection technique is used to calculate fluid and particle velocities, as detailed in \citet{rusche2002computational, weller2002derivation}. 
Pressure-velocity coupling was solved using the PIMPLE method, combining the well-established SIMPLE and PISO algorithms \citep{ferziger2019computational, versteeg1995finite}. The solution procedure is detailed in Appendix~\ref{ap:procedure}.

To compute the forces between phases, only the drag force, calculated from the Gidaspow Wen-Yu/Ergun model \citep{gidaspow1994multiphase}, was considered: 
\begin{equation}
    \textbf{F}_1 = \phi_1 0.75Cd(\text{Re}_p)\rho_2\frac{\nu}{d^2}(\textbf{U}_1 - \textbf{U}_2),
\end{equation}
where $\text{Cd}(\text{Re}_p)$ depends on the model selected, when particle concentration is lower than 0.2 it use the Gidaspow - Wen Yu model, otherwise it uses the Gidaspow-Ergun model:
\[
     \text{Cd}(\text{Re}_p) =  
\begin{cases}
    24\left(1 + 0.15 \text{Re}_{p,c}^{0.687}\right) \phi_2^{-2.65},& \text{if } \phi\leq 0.2\\
    \frac{4}{3}\left(150\frac{\phi_1}{\phi_2} + 1.75\text{Re}_p\right)\phi_2^{-2.65},              & \text{otherwise.}
\end{cases}
\]
Here $\text{Re}_{p,c} = \phi_2\text{Re}_p$, with $\text{Re}_p = \frac{|\textbf{U}_1 - \textbf{U}_2|d}{\nu_2}$. 
Although other forces, such as lift and virtual mass, might be relevant, we assume that drag is the dominant force under our operating conditions, which include low-concentration suspensions, small particle sizes, and laminar flow. This is because, at low Stokes numbers (here $St\approx \rho_sd^2\bar{U}/(\mu L)$) and with a density ratio greater than 1, inertial forces lose significance even in turbulent flows \citep{zhang2021fluid, mortimer2017particle}. Moreover, \citet{volpe2019sorting} indicates that when the particle-channel Reynolds number (here $Re_{pc} = Re(d/B)^2$) is much less than 1 ($Re_{pc} \sim 10^{-5}$ in our simulations), drag primarily drives particle migration.  In the sediment zone (where the thickness is less than 1 mm and the concentration is higher than $\phi_0$), it is assumed for simplicity that gravitational forces become dominant due to the increased density relative to the suspension zone. Additionally, as the concentration increases, the solid pressure term in the momentum equation becomes significant. However, for higher concentrations where the sediment layer is thicker, additional forces, such as collision forces, should be considered.

On the other hand, the shear stress tensors are as follows:
\begin{align}
    \bm{\tau}_i = \mu_i\left[ \left(\nabla\textbf{U}_i + \nabla\textbf{U}_i^T\right) - \frac{2}{3}(\nabla\cdot\textbf{U}_i)I\right]
\end{align}
where $\mu_i = \rho_i\nu_i$ , with $\nu_i$ the kinematic viscosity. To determine an effective viscosity for the solid, it is established that the viscosities must satisfy the following relationship: $\mu_{mix} = \phi_1\mu_1 + \phi_2\mu_2$, where $\mu_{mix}$ corresponds to the viscosity of the mixture calculated as \citep{krieger1959mechanism}:
\begin{align}
    \mu_{mix} = \mu_2\left(1 - \frac{\phi}{\phi_{max}}\right)^{-\zeta\phi_{max}}
\end{align}
where $\phi_{max} = 0.6$ and $\zeta = 2.5$. This equation is relevant for $\phi > 0.1$ where there is a collision between particles.
Finally, the equation to determine the particle concentration is given by:
\begin{align}\label{eq:part_calc}
    \frac{\partial \phi_1}{\partial t} + \nabla\cdot \left(\phi_1 \textbf{U}_1\right)  = 0.
\end{align}
In OpenFOAM this equation is solved explicitly by the MULES algorithm \citep{damian2014extended}, which limits convective flux to allow the variable to remain within a range given by the user.

Numerically, a second-order linearUpwind scheme \citep{warming1976upwind} was used for convection, employing upwind interpolation weights with an explicit correction based on the local cell gradient \citep{greenshields2022notes}. Time discretization was performed using the Crank-Nicholson scheme \citep{crank1947practical}, achieving second-order accuracy by evaluating terms at the midpoint between time steps \citep{greenshields2022notes}. An adaptive time step was implemented to maintain a Courant number near 0.9, minimizing numerical viscosity and ensuring stability of the explicit MULES scheme. Furthermore, the presence of physical instabilities required transient resolution,making an adaptive time step appropriate for modeling this system.

%A summary of the steps of the algorithm that solves these equations can be found in the~\ref{appendix:solver}.

\subsection{Boundary conditions}\label{sec:BCs}

The following boundary conditions have been considered, with reference to the particular methods used in OpenFOAM:
\begin{itemize}
    \item $\mathbf{Inlet}$: A fixed value (Dirichlet boundary condition) was applied for both particle concentration and velocities. The `fixedFluxPressure' condition was used for the pressure, to satisfy both the momentum equation and the boundary conditions of velocities.
    
    \item $\mathbf{Outlet}$: For the particle concentration at the outlet, we utilized the `inletOutlet' condition, which incorporates Neumann-type boundary condition for the outflow and Dirichlet for the inflow. Additionally, the boundary condition `pressureInletOutletVelocity' was applied to manage velocity, which corresponds to a mixed boundary condition where for outflow, a zero gradient condition was utilized, whereas for inflow, the velocity was derived from the internal cell value normal to the patch. On the other hand, a `prghPressure' boundary condition was adopted for the pressure. This condition calculates the pressure at the boundary, excluding the hydrostatic effect, represented by $p_{rgh} = p - \rho g(h - h_{ref})$. 
    
    \item $\mathbf{Hopper}$: In hopper zone,  on a height $H_{Hop}$, the return of excess particles is avoided by imposing that the concentration $\phi$ is equal to $\phi_0$ (Dirichlet condition) in this zone, thus preventing cell contamination from the hopper.  This effectively renders the hopper as a particle sink, thus providing a focus on the dynamics inside the cell rather than on the particular features of the inlet and outlet.
    
    \item $\mathbf{Walls}$: For the walls, a `non-slip condition was chosen for the liquid phase velocity and a free-slip condition for particle phase velocity. `fixedFluxPressure' was utilized for the pressure boundary condition. The particle concentration used a `zeroGradient' condition, which means $\frac{\pa\phi}{\pa n} = 0 $ , where  $n$ corresponds to the normal direction to the wall. 

\end{itemize}

 Since the solid phase is treated as a continuum, the selection of the appropriate boundary condition is not straightforward and may depend on the specific problem being addressed.
Solid particles interact with the wall and can also collide with one another, rotate, or be displaced by the shear stress of the fluid. Consequently, a non-slip condition is not the most suitable choice because the particles do not adhere to the bottom; instead, they tend to slip and may undergo shear-induced mixing with the upper part of the particle layer \citep{benyahia2005evaluation, johnson1987frictional}.
The particles are likely to slip at the bottom, which can be influenced by factors such as roughness of the wall, particle collisions, and friction between the particles and the wall. For this reason, various models have been developed, including the Johnson-Jackson model \citep{johnson1987frictional}, to describe partial slip conditions. However, these models often require the specification of coefficients that may not be readily available from experimental data.
In cases where smooth, non-adhesive walls and hard particles are assumed, the 'free-slip' condition has been found to yield good agreement with experimental results, resembling the results obtained with the partial slip model when using small friction coefficients ( \citet{benyahia2005evaluation}). For this reason, we assume a smooth wall, with hard and non-adhesive particles.

However, the free-slip model, which assumes no wall friction, may become inadequate at high particle concentrations. Under such conditions, deviations in the near-wall profile can occur. Furthermore, the inlet concentration significantly influences sediment concentration and wall-particle interactions. As inlet concentration increases, resulting in higher sediment concentration near the wall, a partial-slip approach, such as the Johnson-Jackson model \citep{johnson1987frictional}, may be necessary to accurately capture these interactions.

Conversely, when the smooth wall assumption is invalid, roughness must be considered. Even in laminar flow, surface roughness can induce friction effects similar to those in pipe flows \cite{liu2019roughness}, primarily impacting the sediment bed by altering particle profiles and accumulation, akin to turbulent conditions. Roughness can reduce streamwise velocity and increase wall-normal velocity, as seen in turbulent flows \cite{milici2015influence, luo2019effects}, enhancing particle resuspension. Furthermore, adhesive forces may lead to preferential particle accumulation, forming a thicker sediment bed. In industrial applications, low surface roughness is preferred to minimize mixing between sediment, suspension, and water layers and to promote self-cleansing.

\section{Initial conditions and convergence to a statistically steady state}\label{sec:convergence}

\begin{figure*}[!ht]
\centering
\includegraphics[width=16.2cm]{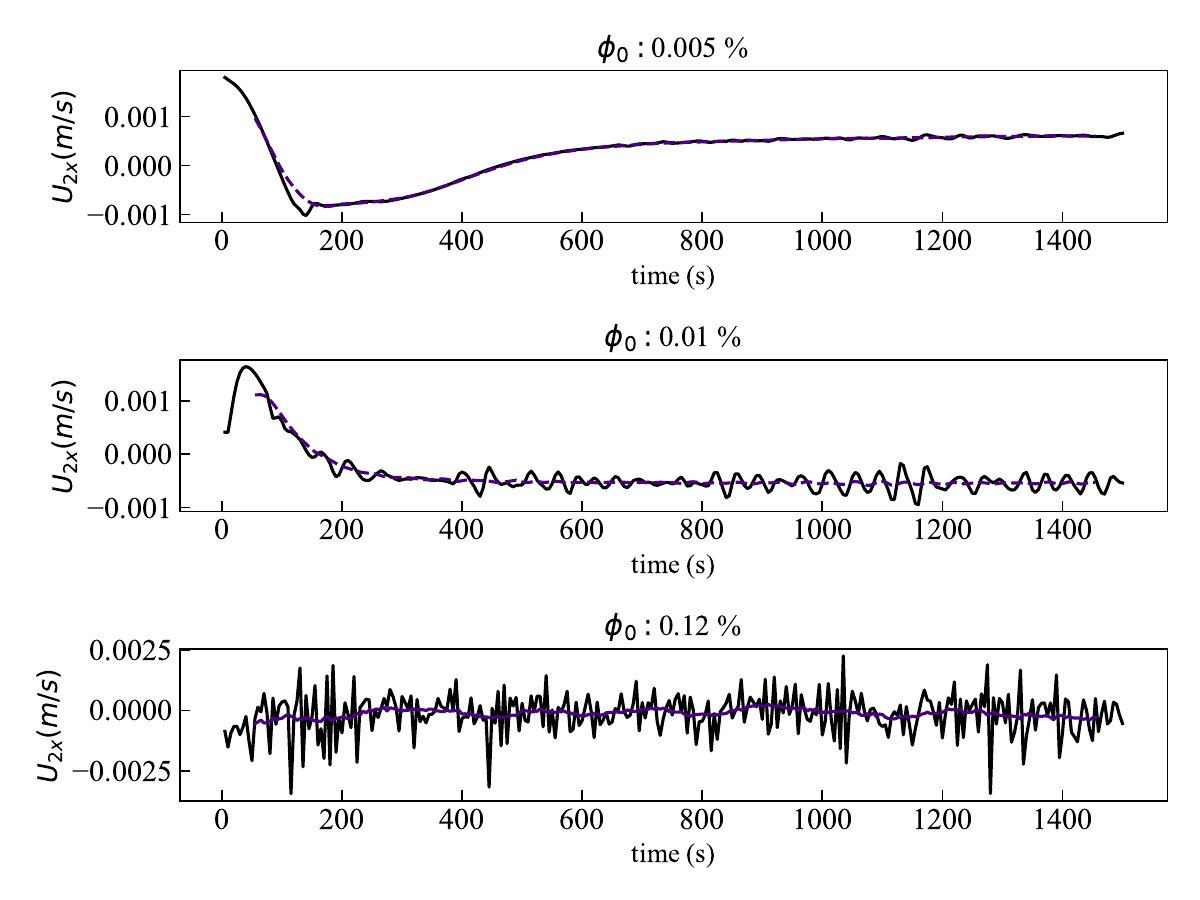}
\caption{Example of velocity field convergence to steady-state behavior at a fixed point (0.005, 0.27) for three different particle concentrations. Here $U_{2x}$ is the longitudinal water velocity.}\label{f:convergence}
\end{figure*}

To characterize the flow in what would correspond to continuous operation conditions, simulations need to be carried out until they reach a steady state. This can be either a stationary state or a fluctuating but statistically steady state, which is achieved over several flow cycles once stationary mean values have been reached. 
Examples of this convergence to a steady state are shown for three particle concentrations in Figure~\ref{f:convergence}, which shows the instantaneous value (solid lines) of the longitudinal velocity of the liquid $U_{2x}$ at a specific point (indicated by white dots in Figure~\ref{f:resuspension} below) and its moving average (dashed lines) on a centered window of 100 time units. It is observed that the convergence of the flow values can show different types of variation involving different time scales which, as apparent from the simulations, depend on the development of flow instabilities. For example, for a system that does not show instability (top plot of Figure~\ref{f:convergence}, $\phi_0 : 0.005\%$), the flow evolves from its initial condition to smoothly converge to its final values.  In the presence of instabilities, the flow can oscillate around a value with a clear dominant frequency (middle plot, $\phi_0: 0.01\%$) or with a strongly fluctuating behavior (bottom plot $\phi_0 : 0.12\%$). In the first two cases (top and middle plots), the convergence of the system to an average value is clearly observed in the moving average. The small oscillations detected at $\phi_0 = 0.01\%$ (solid line, middle plot) suggest that the system is near the onset of an instability. For the third case, with a much higher particle concentration ($\phi_0 = 0.12\%$, bottom plot), the moving average shows slight variations indicating that the system has broad spectrum fluctuations reaching low frequencies. Still, the system does not show a clear long-term trend and thus appears to be sufficiently close to converge to a well-defined average. 
We define as a steady state the full velocity and concentration fields corresponding to the average of the last 20 snapshots, with a time difference of $\Delta t = 5$ s in between.
Residual variations of the point-wise variables (as in the moving average in the bottom plot of Figure~\ref{f:convergence}) will, after initial transients have passed, have a negligible effect on integral measures such as the local TRE and the positions of the interface used to characterize the settler.

\begin{figure*}[!ht]
\centering
\includegraphics[width=16.2cm]{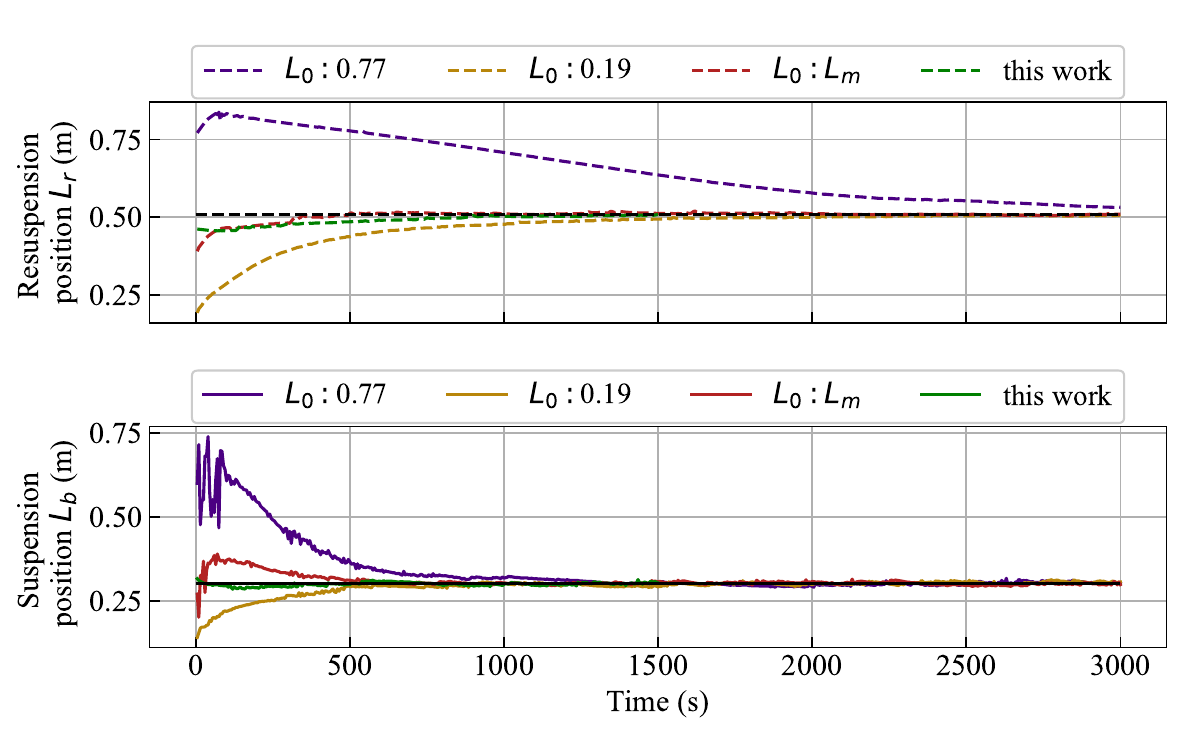}
\caption{Effect of initial conditions on reaching the steady state, where the suspension position ($L_b$) is $L_0$ at $t = 0$ with a uniform particle concentration $\phi_0$. $L_m$ is calculated using eq~\ref{eq:LmDef} whose value for this case equals 0.389 m.}\label{f:initial_cond}
\end{figure*}

The choice of initial condition is crucial to reduce transients and accelerate convergence in the various cases. 
For the first case, a homogeneous concentration up to a certain height $L_0$ was defined inside the cell, with zero velocity for both phases. 
This results in a long adjustment time until the steady state is reached.  
The evolution of these transients can also be studied by following the position of the interfaces, both of the bulk suspension $L_b$ and of the resuspension $L_r$, as shown in Figure \ref{f:initial_cond}. 
It is observed, as expected, that the further away the initial condition is from the steady state, the more time is needed to reach that steady state (see $L_0 = 0.77$ and $L_0 = 0.19$ in Figure \ref{f:initial_cond}). 
Furthermore, when the initial condition is set significantly above the steady-state height, the system exhibits instabilities, leading to oscillations in the interface position at the beginning of the simulation. Conversely, initial conditions below the steady-state height result in smoother convergence.

When the initial condition is close to the steady state, such as for low concentration when $L_0=L_m$, the guess based on the theoretical solution (with zero velocity and a uniform particle concentration up to a height $L_m$, Eq.~\eqref{eq:LmDef}), or the solution of a simulation with a close particle concentration, then the system reaches the steady state in a faster way. 
In this work, most cases were run from nearly the lowest particle concentration ($\phi_0=0.005\% $) to the highest, using the steady-state concentration of the previous case as an initial condition for the concentration in next, after normalizing its values to the new concentration. % with the objective of having reference values close to the new steady state.  
In this way, it is possible to reduce the computational time because the initial condition is closer to the steady state. % than starting from a fixed condition with zero velocity or a random initial condition.
We stress that the main advantage of this method %used in this research 
over using the theoretical solution $L_m$ is that the theoretical solution gets increasingly farther away from the final solution as the particle concentration increases (a topic to be discussed further in the results).
% The  , so it will not always be the fastest method to reach the final state faster.

It is important to note that different initial conditions may result in the convergence to distinct steady states, as can be expected in the presence of multiple attractors of the dynamical system. 
Moreover, since our protocol for the choice of initial conditions involves an increase in particle concentration, one could expect to find hysteresis when concentration is decreased.
We performed a few tests to explore these possibilities and found, in all such attempts, that the system ultimately converges to the same steady state, independent of the initial condition.

The computational domain discretization also plays a central role in the accuracy and stability of the simulations. A detailed mesh resolution study is presented in Appendix~\ref{appendix:mesh}.

\section{Results}

We begin in \S \ref{sec:resultsConcentration}  by presenting the simulations for different inlet concentrations with a medium SOR (\SI{0.036}{mm/s}) and an angle of \SI{55}{\degree}, %aiming to identify a case with 
which provides a moderate degree of resuspension, %. This setup was chosen because it provides 
and a smooth transition of instability regimes. 
The effects of SOR and angle of inclination are discussed independently in the following subsections.

\subsection{Particle concentration effect}\label{sec:resultsConcentration}

\begin{figure*}[!ht]
\centering
\includegraphics[width=17.2cm]{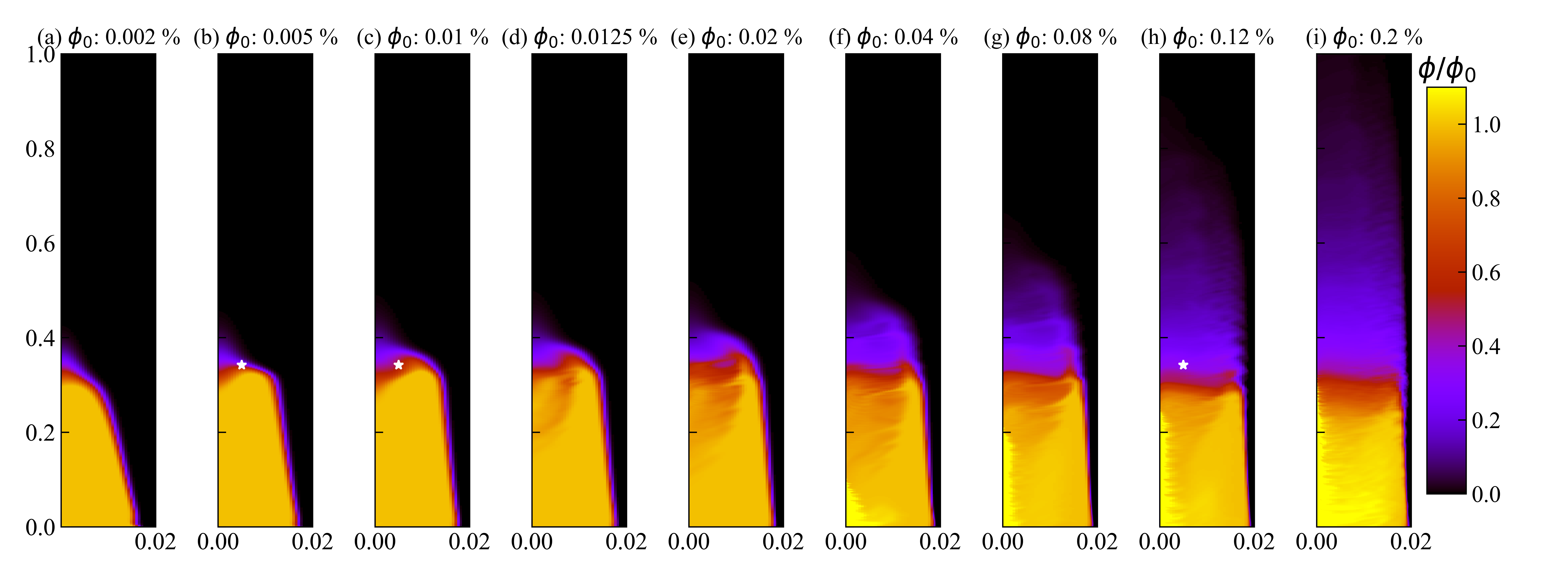}
\caption{Particle concentration profile for 9 different inlet particle concentrations. Orange zone corresponds to bulk suspension and purple zone corresponds to resuspended particles. Here SOR: \SI{0.036}{m/s}, $\theta$ : \SI{55}{\degree} and Re: 30. The white dot corresponds to the point where the convergence test was performed (Figure~\ref{f:convergence}).}\label{f:pc_effect}. 
%where R0: (a-c), R1: (d,e), R2: (f,g) and R3 (h,i). 
\end{figure*}

Figure \ref{f:pc_effect} shows the mean particle in the 2D profile for 9 distinct particle concentrations, normalized by input concentration. An observable trend emerges: with the rise in concentration, there is a corresponding increase in particle resuspension (depicted by the purple zone). Consequently, this leads to a decrease in cell efficiency due to the increase in the number of particles reporting to the overflow, diverging from the theoretical value of inclined plate designs such as \citet{Yao70JotWPCF}.

\subsubsection{Physical description of the flow development}
In the simulations, we observe the gradual emergence of instabilities that intensify with increasing particle concentration. 
For the lowest concentrations, the flow settles into a state without time dependence. 
We will refer to this as Regime 0.
Flow fluctuations become noticeable in the steady-state concentration profiles when the inlet concentration exceeds $\phi_0 = 0.01\%$. 
At this point the effects of time variation remain confined to a small region below the bulk interface position, as evidenced by the darker regions (brick color) of reduced concentration within the bulk suspension in Figure \ref{f:pc_effect} (d-e). 
Until this concentration level, which includes Regime 1 to be described below, there is minimal to no resuspension towards the upper region of clarified fluid. 
As the particle concentration increases to $\phi_0 = 0.04\%$ and beyond, the region of particle resuspension increases until the resuspended particles cover almost the entire upper section of the settler, as illustrated in Figure \ref{f:pc_effect} (i). 
We distinguish here two further regimes: Regime 2 in which the density of resuspended particles decreases in a staircase structure with nearly discrete changes, as shown in Figure \ref{f:pc_effect} (f, g), and Regime 3 in which the resuspension becomes diffuse (Figure \ref{f:pc_effect} (h, i)). 

\begin{figure*}[!ht]
\centering
\includegraphics[width=15.2cm]{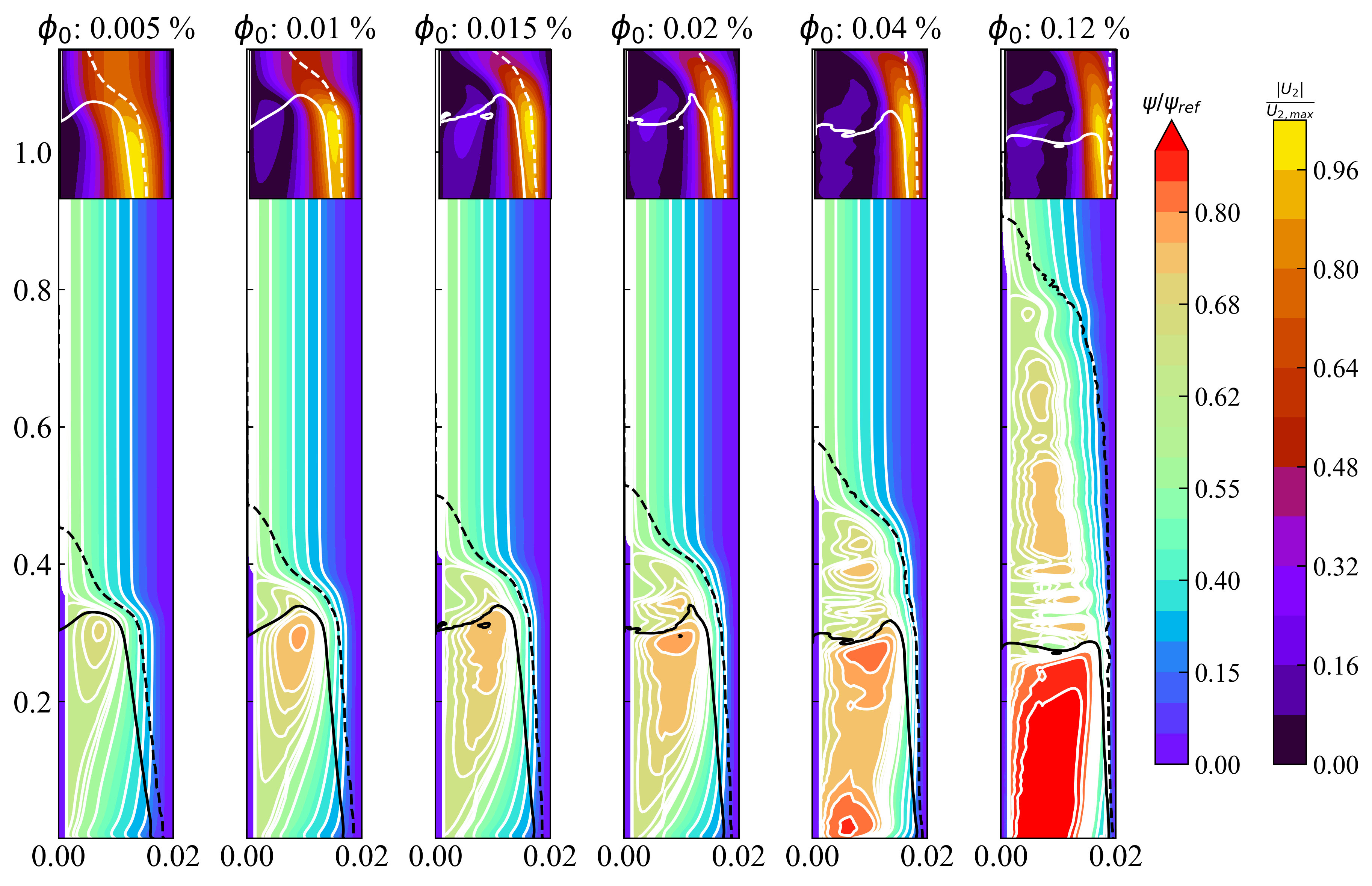}
\caption{Streamfunction field ($\psi$) normalized by a reference value ($\psi = \num{5e-3}$ \SI{}{m^2/s}) for six different particle concentrations, SOR: 0.036 m/s, $\theta=\SI{55}{\degree}$, $\text{Re}= 30$ (same cases as Figure~\ref{f:pc_effect}). The thick solid line on each plot corresponds to bulk suspension ($\phi=0.8\phi_0$) and the dashed lines correspond to resuspended particles ($\phi=0.005\phi_0$). The inset at top of each figure corresponds to the normalized magnitude of velocity field of the fluid phase $\textbf{U}_2$, the yellow zone representing the zone with the highest velocity within the whole domain. The clear fluid zone is at the right in each figure.  %solid line corresponds to and dashed line corresponds to. dotted gray line is where $\omega = 0$.
}\label{f:stream}
\end{figure*}

\subsubsection*{Initial regimes of destabilization (from regime 0 to 1): Separation of the clear fluid from the suspension}

\begin{figure*}[!ht]
\centering
\includegraphics[width=15cm]{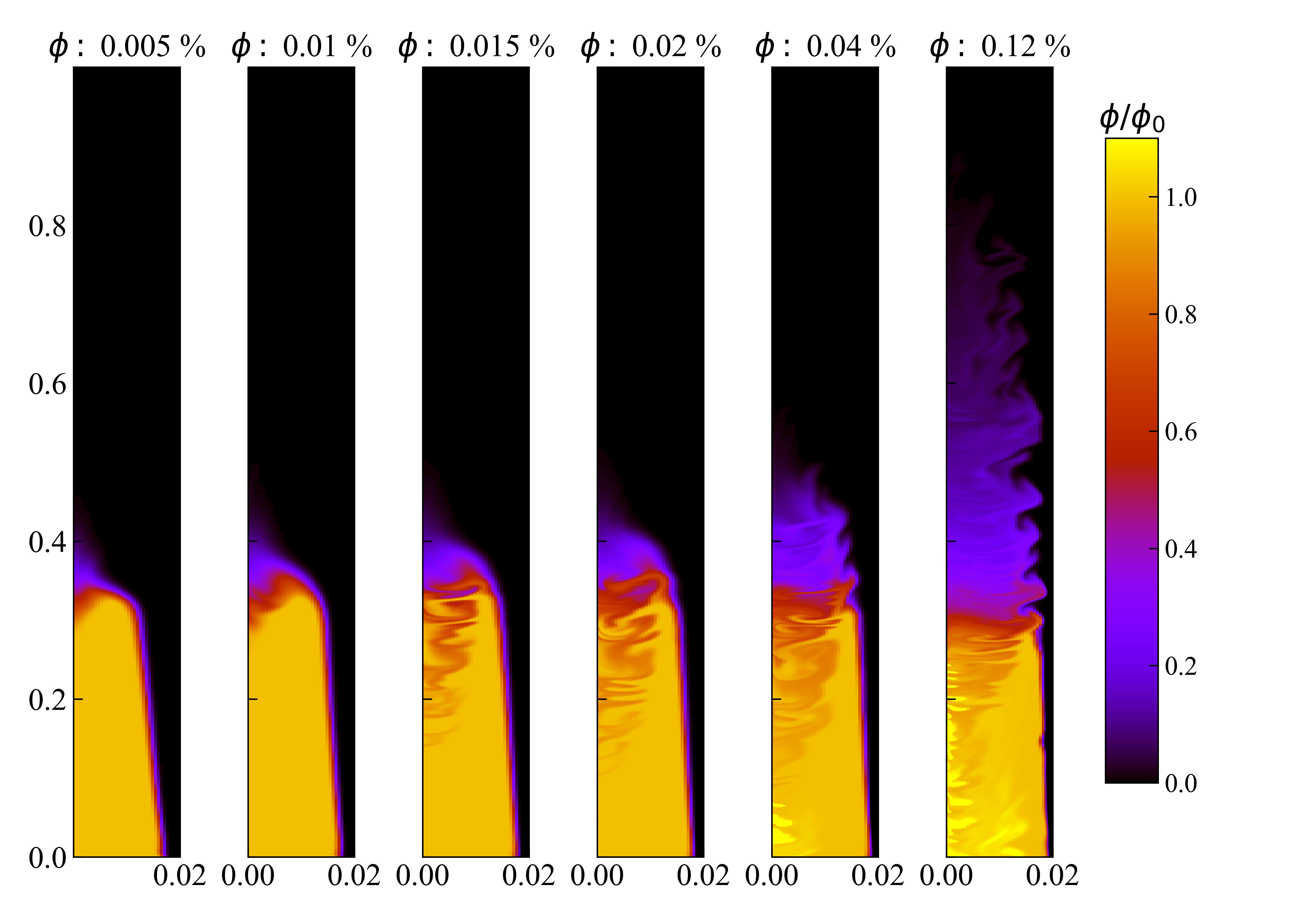}
\caption{ Time-evolving simulation of the 2D particle concentration profile for various inlet concentrations (multimedia available online).}\label{f:matsup}
\end{figure*}

Changes in the configuration of the (statistically) steady state through the different regimes are detailed in Figure \ref{f:stream}, which shows contours of the stream function of the (volumetric) mean velocity $\textbf{V}$ and isolines that indicate regions occupied by the bulk of the suspension (solid black line, $\phi=0.8\phi_0$) and particle resuspension (dashed line, $\phi=0.005\phi_0$). 
The first three subfigures (from left to right) show the transition from a stable stationary system to a time-dependent but only mildly unstable one. 
For $\phi_0=0.005\%$, the stationary flow has one recirculation zone located around the tip of the suspension region (bounded by the solid black line), just below a single stagnation point on the lower wall (left side). 
In the configuration at $\phi_0 = 0.01\%$, the top of the recirculation zone is stretched and tilted to the left, while the steady mean flow detaches from the wall. 
In the time-developing simulations (see Figure\ref{f:matsup}, multimedia available online), these features are seen to correspond to the release of small jets of particles from the tip of the suspension, similar to what \citet{laux1997computer} have referred to as a `fountain of particles.   
The topological configuration of the streamlines remains unchanged up to an initial concentration $\phi_0=\SI{0.015}{\%}$.

We identify this time-dependent regime with a single recirculation zone away from the wall as Regime 1. 
At this point, the clear fluid separates from the suspension and the small recirculation zone that appears remains trapped between the suspension and the wall. 
Unlike the incipient case with $\phi_0=0.01\%$, for $\phi_0=0.0125\%$ (and $\phi_0=0.015\%$) the jets are strong enough to lead to the formation of overhanging zones which are gravitationally unstable and break down in plumes like those of the Rayleigh-Taylor instability, but with shear. 
These plumes induce mixing, near the interface, of clear water entrained into the bulk suspension. 
These unstable overhanging zones are at the origin of the darker regions (brick color) of the reduced concentration within the bulk suspension in Figure \ref{f:pc_effect}(d-e) mentioned above. 
At the high concentration end of this range of $\phi_0$, \emph{i.e.}, for $\phi_0=0.02\%$, new features begin to appear in the streamlines of Figure \ref{f:stream}.

\subsubsection*{Regime 2: Detachment and suspension release}

Figure \ref{f:stream} shows that the recirculation zone around the tip of the suspension splits in additional vortices when the inlet concentration increases to $\phi_0=0.02\%$, leading to a new recirculation region. 
Some more recirculation zones are visible for $\phi_0=0.04\%$, most of which extend upward along the settler and are tightly contained within the resuspension region (dashed line). 
The location of these upper recirculation zones coincide with the steps (or nearly discrete changes) in the mentioned staircase structure of the density of resuspended particle concentration in Figure \ref{f:pc_effect}(f) (also visible in Fig. \ref{f:pc_effect}(g)). 
These results indicate that the growth of the resuspension region is associated with enlargement of the clear fluid separation region and the detachment of the recirculation zones from around the suspension interface. 
These recirculation zones break up the suspension, releasing particles into a resuspension region that extends up to around the point where the clear fluid reattaches to the wall. 
In fact, it is noted that the resuspension region (dashed line) closely follows the streamlines that envelope the recirculation zones, which is a distinguishing feature of Regime 2. 

Another characteristic that becomes noticeable in this regime is the appearance of a distinct recirculation zone near the bottom of the suspension region (lower left corner of $\phi_0=0.04\%$ in Figure \ref{f:stream}); this corresponds to a marked increase in particle concentration within the suspension (bright yellow corner in Figure \ref{f:pc_effect}(f)). 
%this bottom and top recirculation zones within the suspension 
When the inlet concentration increases further to $\phi_0=0.12\%$, this bottom recirculation zone has merged with the one at the tip of the suspension into a single open cell engulfing the bulk of the suspension. 
This induces a further increase in particle concentration within the suspension (enlarged bright yellow regions in Figure \ref{f:pc_effect}), this time with particles transported from the sediment region near the underflow outlet of the settler. 
Furthermore, since the difference of the values of the stream function at any two points gives the total flux (of incompressible flow) between those points, the increasingly red color in Figure \ref{f:stream} relates to a marked increase in flow velocity. 
The location where the velocity of the fluid phase $\textbf{U}_2$ is maximum can be seen in the insets of Figure \ref{f:stream}, which show its normalized absolute value $|\textbf{U}_2|/\max(|\textbf{U}_2|)$ around the tip of the suspension. 
The maximum velocity is in all cases located next to the tip of the suspension, close to the boundary between the suspension and clear water layer. 
The increase in flow velocity around the interface between the clear fluid and the suspension is fully aligned with the sequential regimes of onset and enlargement of flow separation described so far. 
However, as the inlet concentration reaches $\phi_0=\SI{0.12}{\%}$, other effects become important. 

\subsubsection*{Regime 3: Erosion of the clear-fluid\textendash suspension interface}

The plot on the right of Figure \ref{f:stream} shows several recirculation zones within a substantially enlarged resuspension region, compared to previous cases. 
Moving upward from the top of the suspension interface (black line), the width of the recirculation zones decreases considerably, while the width of the resuspension region remains largely constant. 
%This is unlike the previous Regime 2 in which the shape of the resuspension region closely follows the streamline that envelopes the recirculation zones. 
This indicates that the resuspension region contains particles that do not originate in the recirculation zones but are extracted at the interface between the clear fluid and the suspension. 
We distinguish two ways in which this erosion of the clear-fluid\textendash suspension interface takes place.

First, the observed resuspension can be partly linked to shear instabilities, driven by %a reduction in the thickness of clear water and (me parece que esta reduccion no favorece las shear instabilities)
an increase in the maximum velocity of the clear water layer as $\phi$ increases (this velocity increase is explored further in \S \ref{sec:discussion}). This creates growing interfacial waves at the clear-fluid\textendash suspension interface, propelling a portion of the particles towards the outflow. 
This is the mechanism that has been studied most extensively in the literature to explain the loss of efficiency in inclined settlers \citep{davis1983wave,davis1985sedimentation,shaqfeh1987Stability,borhan1988sedimentation}. 
Indications of the development of a shear instability at the clear-fluid\textendash suspension interface can be found for the largest concentration in Figure \ref{f:pc_effect}(i), which shows oscillations in the density of particle concentration near the interface (brick color indentations into the suspension area near $(x,y)=(0.2,0.02)$). 
For the other case with $\phi_0=0.12$ in Figure \ref{f:pc_effect}(h), a shear instability is only apparent in the wavy shape of the edge of the resuspension region (and even this may potentially be related to the recirculation zones visible in Figure \ref{f:stream}).
Being largely developed above the suspension, this mechanism does not fully explain the particle erosion that originates in the resuspension region. 

The second mechanism of erosion of the suspension interface relates 
not only to the magnitude of maximum velocity of the fluid, but also to its position within the system. 
As seen in the insets of Figure \ref{f:stream}, the highest velocities occur around the clear water layer, next to the suspension interface, encouraged by the rapid ascent of water due to the Boycott effect. 
However, as the velocity increases with increasing $\phi_0$, there is a strong reduction in the thickness of the clear water layer, to the point at which the maximum velocity overlaps the clear-water\textendash suspension interface. 
Indeed, when the concentration reaches $\phi_0 = 0.12\%$, the maximum velocity is found within the suspension itself, which can significantly increase the resuspension of particles. Figure~\ref{f:stream} (inset) captures this phenomenon, showing the shift to the right of the particle interface that results in the highest velocities and shear rates occurring within the suspension. This observation aligns with the increased resuspension observed within the system and highlights the critical role of the velocity and thickness of the clear fluid layer under varying particle concentrations.
A scaling analysis and the relationship between thickness, maximum velocity, and particle concentration will be discussed further in Section~\ref{sec:discussion}.

%\begin{figure}[!ht]
%\centering
%\includegraphics[width=8cm,keepaspectratio]{interphase_n.eps}
%\caption{The suspension position ($L_b$) and resuspension position ($L_r$) as a function of inlet particle concentration are shown, where the black dashed line represents the theoretical line $L_m$ obtained from equation~\ref{eq:LmDef}.}\label{f:resuspension}
%\end{figure}

%\begin{figure}[!ht]
%\centering
%         \includegraphics[width=8cm]{TRE.eps}
%\caption{The effect of $\phi_0$ on local efficiency (TRE) is shown for three different positions along the cell. In both figures, $\text{Re} = 30$, $\text{SOR} = \SI{0.036}{m/s}$, and $\theta: \SI{55}{\degree}$.}\label{f:resuspension}
%\end{figure}

\begin{figure*}[!ht]
\centering
    \begin{tabular}{cc}
        (a) & (b) \\
         \includegraphics[width=8cm,keepaspectratio]{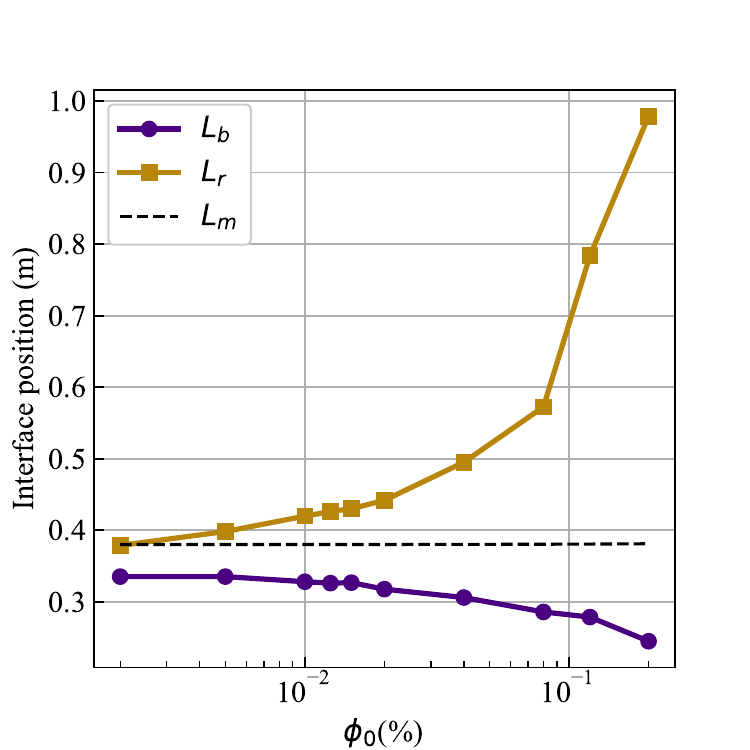}&
         \includegraphics[width=8cm]{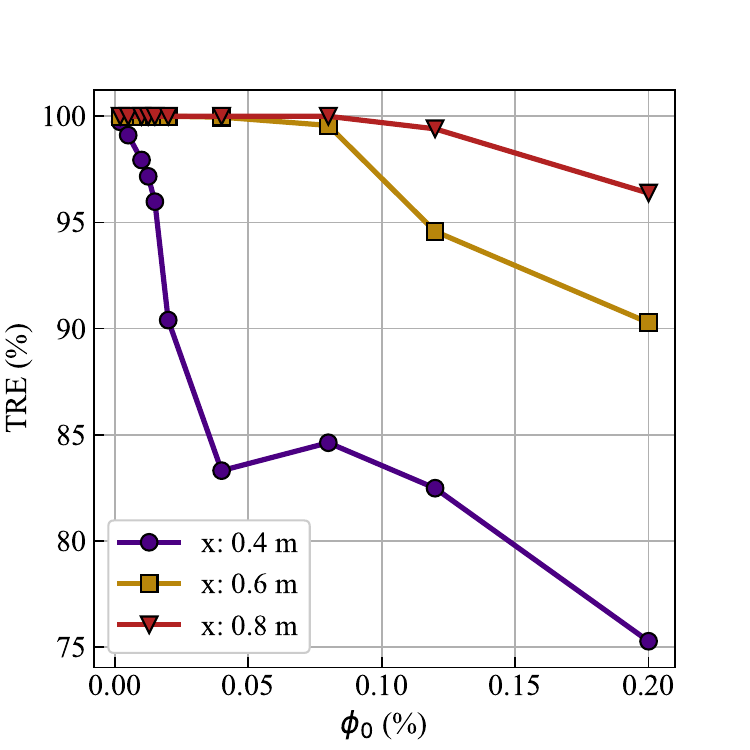}
    \end{tabular}
\caption{On the left, the suspension position ($L_b$) and resuspension position ($L_r$) as a function of inlet particle concentration are shown, where the black dashed line represents the theoretical line $L_m$ obtained from equation~\ref{eq:LmDef}. On the right, the effect of $\phi_0$ on local efficiency (TRE) is shown for three different positions along the cell. In both figures, $\text{Re} = 30$, $\text{SOR} = \SI{0.036}{m/s}$, and $\theta: \SI{55}{\degree}$.}\label{f:resuspension}
\end{figure*}

\subsubsection{Performance quantifications}
The increase in resuspension, due to the increase in concentration and its effect on the bulk suspension, is shown in Figure~\ref{f:resuspension} (left), for the present values of $\text{SOR}= \SI{0.036}{mm/s}$ and tilt angle $\theta=\SI{55}{\degree}$. The figure compares the interface position as predicted by theory and observed in simulation. In all cases, the theoretical line is observed to lie between the positions of the resuspension and the bulk suspension. In addition, the resuspension position ($L_r$) increases significantly, reaching nearly four times its initial value. This change is also reflected in the decrease in efficiency within the settler. 
The loss in efficiency is evident throughout the cell, as Figure~\ref{f:resuspension} (right) demonstrates how the local efficiency decreases at three different positions $x=0.4,0.6,\SI{0.8}{m}$ within the settling plates, showing a sharp drop at all three positions for concentrations higher than $0.08\%$. 
This is a concentration within Regime 2, confirming that the dynamics of the system is strongly affected by the onset and development of instabilities, which in turn depend on the concentration of particles entering the system

\subsection{Surface overflow rate (SOR) effect}

\begin{figure*}[!ht]
\centering
\includegraphics[width=14.2cm]{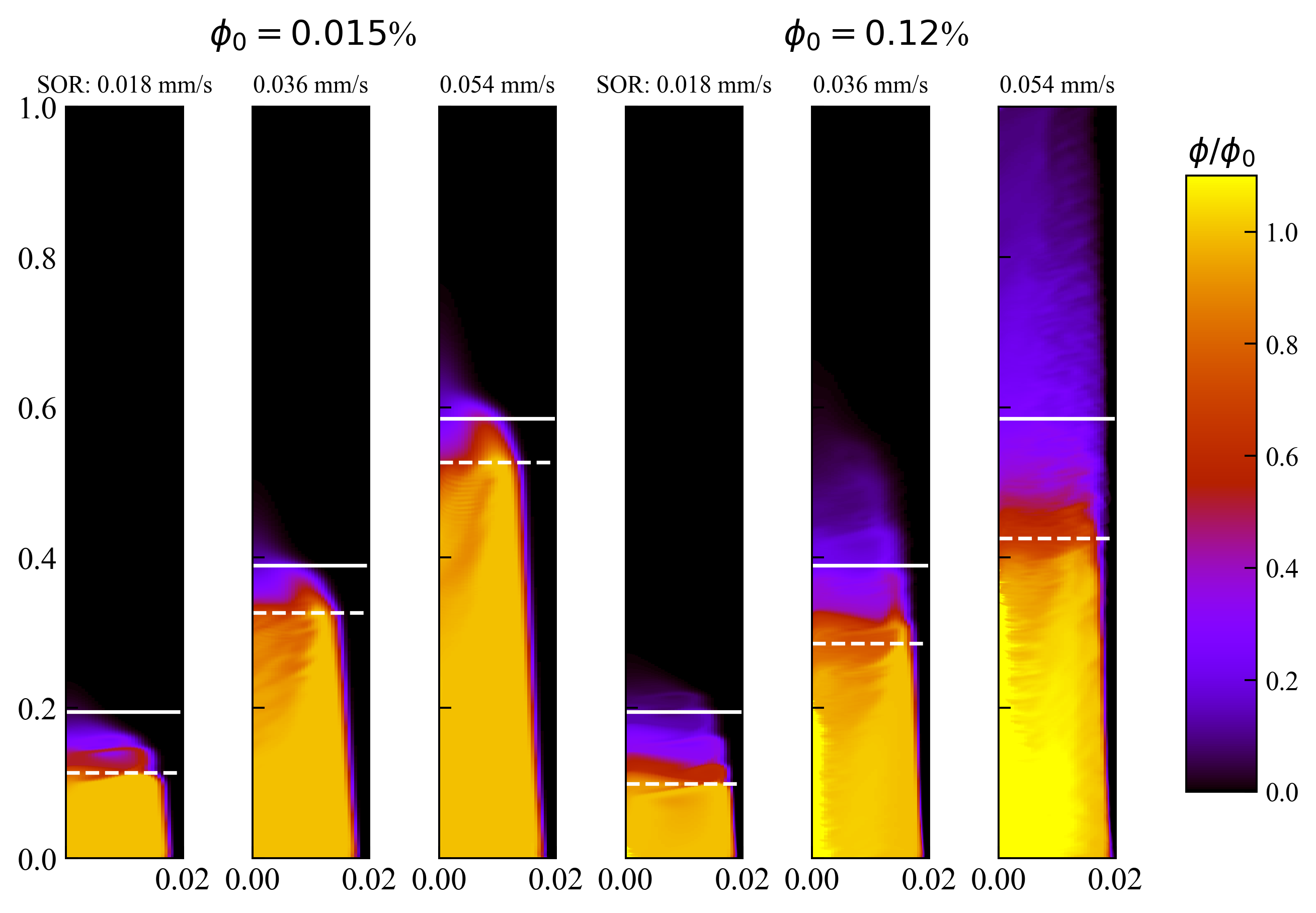}
\caption{Effect of SOR on the particle concentration profile, where the orange zone corresponds to suspension and the purple zone corresponds to resuspension. Here Re = 15-45, $\theta: \SI{55}{\degree}$, and $\phi_0: 0.12\%$. The solid line represents the theoretical line ($L_m$) calculated by Eq.~\eqref{eq:LmDef}, and the dashed line represents the bulk position ($L_b$).}\label{f:sor_prof}
\end{figure*}

The spatial particle concentration profiles for three different surface overflow rates, $\text{SOR}=0.018,0.036,0.054\SI{}{mm/s}$, at two particle concentrations are illustrated in Figure \ref{f:sor_prof}  ($\phi_0=0.015\%$ and $\phi_0=0.12\%$). This figure also demonstrates that the transitions between the four regimes discussed in Section~\ref{sec:resultsConcentration} are influenced by the SOR value. For a concentration of $0.12\%$ the Regime 3 is observed for a SOR of 0.036 and \SI{0.054}{mm/s}, with different degrees of erosion, leading to a greater expansion of the resuspension zone for the higher SOR (purple zone in figure). Clearly, in both cases the theoretical interface position (solid line) fails to accurately represent the true interface position. In contrast, for the $\phi_0 = \SI{0.015}{\%}$  case, the theoretical position aligns closely with the obtained position, even in the presence of instabilities, which are associated with the transition between Regime 1 and Regime 2.

Furthermore, for $\phi = \SI{0.12}{\%}$ an increase in the area resuspended from the sediment (light yellow zone in Figure~\ref{f:sor_prof}) is observed, probably due to the growth of instabilities in the sediment zone as the SOR increases. This suggests that instabilities originating at the suspension interface can affect the stability of the sediment layer and vice versa, since recirculation of the suspension can also influence sediment entrainment into the suspension layer.

This nonlinear relationship between SOR and settler efficiency is consistent with the findings in \citet{demir1995determination} for a multiplate system, where it is shown that efficiency can decrease by more than $\SI{20}{\%}$ when loading is doubled, for a study of Bentonite clay at 70--\SI{80}{NTU}.

\begin{figure*}[!ht]
\centering
    \begin{tabular}{cc}
        (a) & (b) \\
         \includegraphics[width=8cm,keepaspectratio]{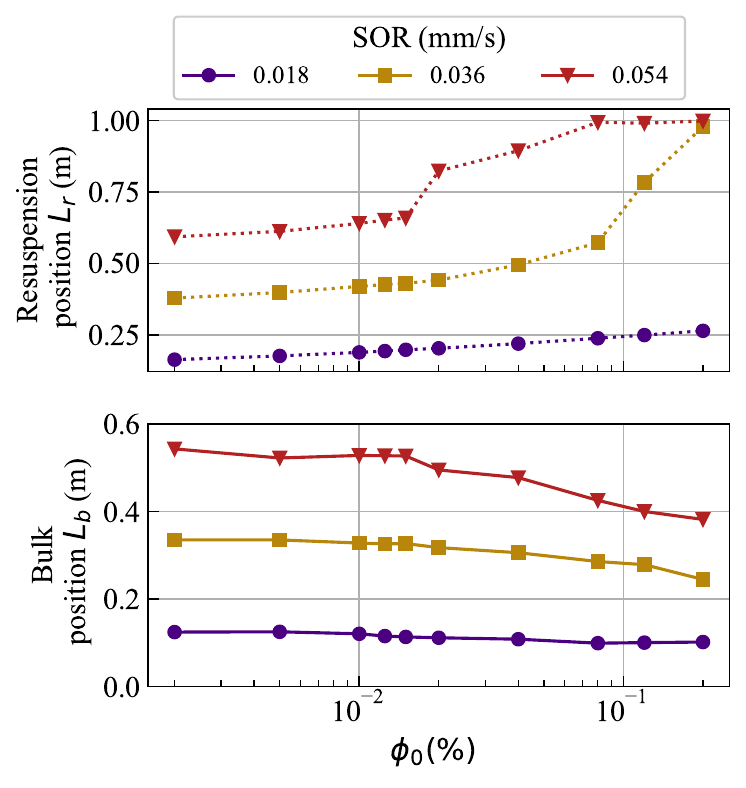}&
         \includegraphics[width=8cm]{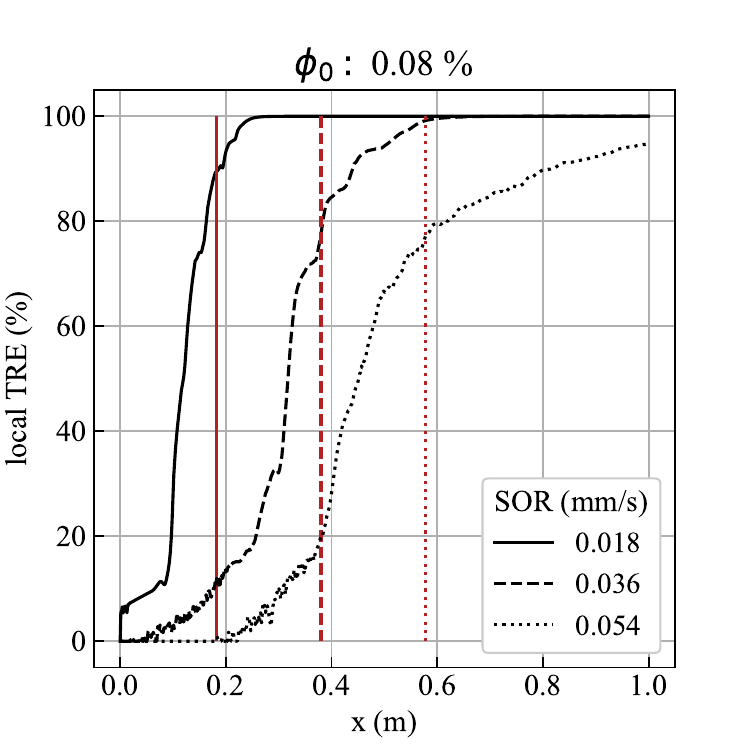}
    \end{tabular}
\caption{ On the left, the effect of SOR on interface positions for three different particle concentrations is shown, where the suspension position ($L_b$) and resuspension position ($L_r$) correspond to the solid and dashed lines, respectively. On the right, the local TRE values along the cell for three different SOR values are shown with $\phi= \SI{0.08}{\%}$. The red vertical lines represent the theoretical value ($L_m$) for each SOR value.}\label{f:resuspensionSOR}
\end{figure*}

The behavior for different concentrations and SOR is shown in Figure~\ref{f:resuspensionSOR} (left). For lower SOR values, the resuspension phase exhibits a gradual increase without reaching the end of the cell within the range studied. At the same particle concentration, the distinctive instabilities observed in higher SOR systems are notably absent in the low SOR case. For low SOR values, Regime 3 is absent within the studied range; instead, the flow is characterized by the dominance of the staircase structure. The lack of instability development in the resuspended region prevents substantial entrainment of particles into the overflow. In fact, for this case, Regime 2 is the dominant regime, appearing at lower concentrations compared to other SOR values and persisting throughout the studied range. In contrast, regimes 0 and 1 have a very limited range and, for certain angles, some of these regimes are not observed.

In contrast, the higher SOR scenario, where the SOR is equal to \SI{0.054}{mm/s}, shows a different behavior: particles reach the overflow zone at a lower concentration compared to the medium and low SOR cases, where the SOR is \SI{0.036}{mm/s} and \SI{0.018}{mm/s}, respectively. This indicates that the transition between different regimes occurs earlier, with Regime 3 being reached at lower $\phi_0$ values (where erosion dominates and shear instabilities are relevant). 
This is further illustrated by the increasing distance between $L_b$ (dashed lines) and the theoretical line $L_m$ (solid lines), as shown in Figure~\ref{f:resuspensionSOR}, which increases with higher SOR values. This growth highlights the pronounced effect of erosion on the suspension, leading to a greater number of particles accumulating above the theoretical line.

According to local efficiency, as depicted in Figure \ref{f:resuspensionSOR} - right,  an increase in the Surface Overflow Rate (SOR) leads to a more elongated profile and a larger region with efficiency below $\SI{100}{\%}$.This profile elongation indicates that a longer plate would be necessary to achieve 100$\%$  efficiency. This observation is particularly evident for $\text{SOR} = \SI{0.054}{mm/s}$, where no position within the cell reaches an efficiency of 100 $\%$. This requirement cannot be obtained from the theoretical model.

\subsection{Inclination angle effect}

\begin{figure*}[!ht]
\centering
    \begin{tabular}{cc}
        (a) & (b) \\
         \includegraphics[width=8.cm]{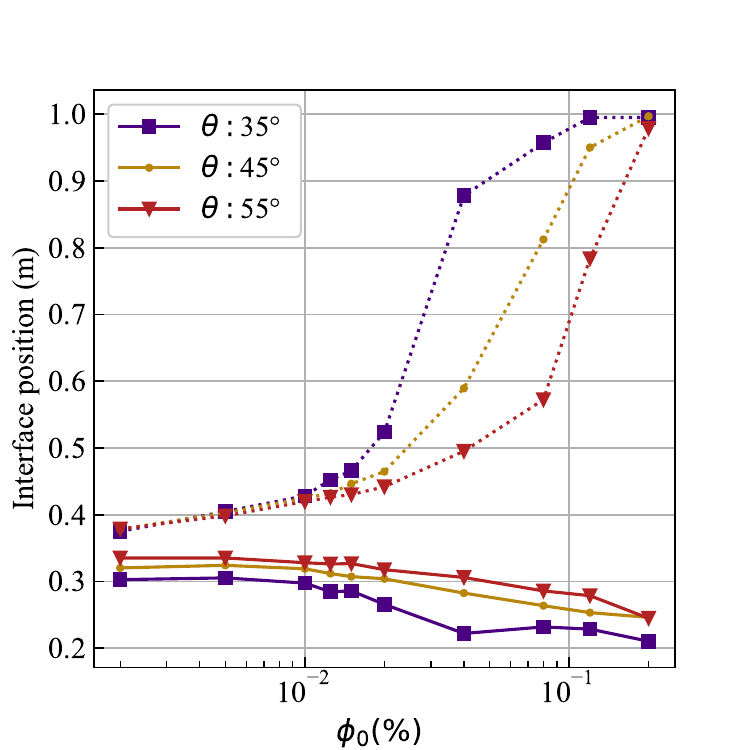}&
         \includegraphics[width=8.cm]{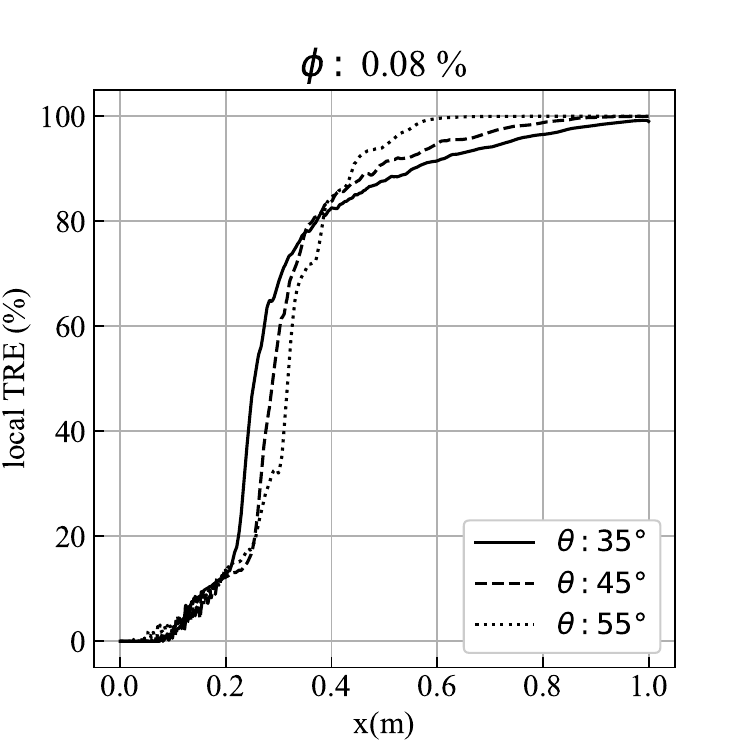}
    \end{tabular}
\caption{Effect of inclination angle ($\theta$) on (a) interface position for three particle concentration, where suspension and resuspension position ($L_b$ and $L_r$) corresponds to the solid and dashed line respectively  and (b) local TRE. Both for SOR: \SI{0.036}{mm/s}.}\label{f:resuspensionAngle}
\end{figure*}

Figure \ref{f:resuspensionAngle} (left) shows that the interface positions behave similarly across three different angles as the particle concentration increases. For the three angles studied, although not all cases exhibit the same regimes (R0-R3), a fixed SOR value tends to result in the same regimes for each angle analyzed. For example, at an SOR of \SI{0.036}{mm/s}, all three angles exhibit an initial stable regime, followed by the onset of recirculation, the development of a staircase structure, and finally the growth of the resuspension zone associated with erosion. The main difference lies in the fact that the transition between regimes occurs at lower concentrations as the cell becomes more vertical (see Figure~\ref{f:summary} for more details). In the most vertical case (\SI{35}{\degree}), the resuspension zone begins to grow for a $\phi_0 = 0.02\%$ (regime 3). On the other hand, for a more horizontal case (55°) this value is around $0.08\%$. This enhancement in stability with increasing angle was also observed experimentally by \citet{borhan1988sedimentation} and later confirmed through numerical simulations by \citet{chang2019three}. The latter points out that when the angle is less than \SI{45}{\degree} the sedimentation efficiency is reduced due to flow instability. In our simulations, this behavior is also reflected in the thickness of the clear water layer, which for the \SI{35}{\degree} case is always less than for the \SI{55}{\degree} case generating a more confined layer and a more elongated velocity profile.

Within the range studied, the effect of varying this parameter on efficiency is less significant compared to the influence of the SOR variable. When comparing the local efficiency profile for the case $\phi_0 = \SI{0.08}{\%}$ the profiles are quite similar when varying the inclination (see Figure \ref{f:resuspensionAngle}, right), which is not the case when varying the SOR. A change in the angle of inclination could affect the efficiency between 1 and $\SI{5}{\%}$ compared to the SOR parameters, where TRE could vary between 10 and $\SI{20}{\%}$.

\subsection{Summary of results}

\begin{figure*}[!ht]
\centering
\includegraphics[width=16.2cm]{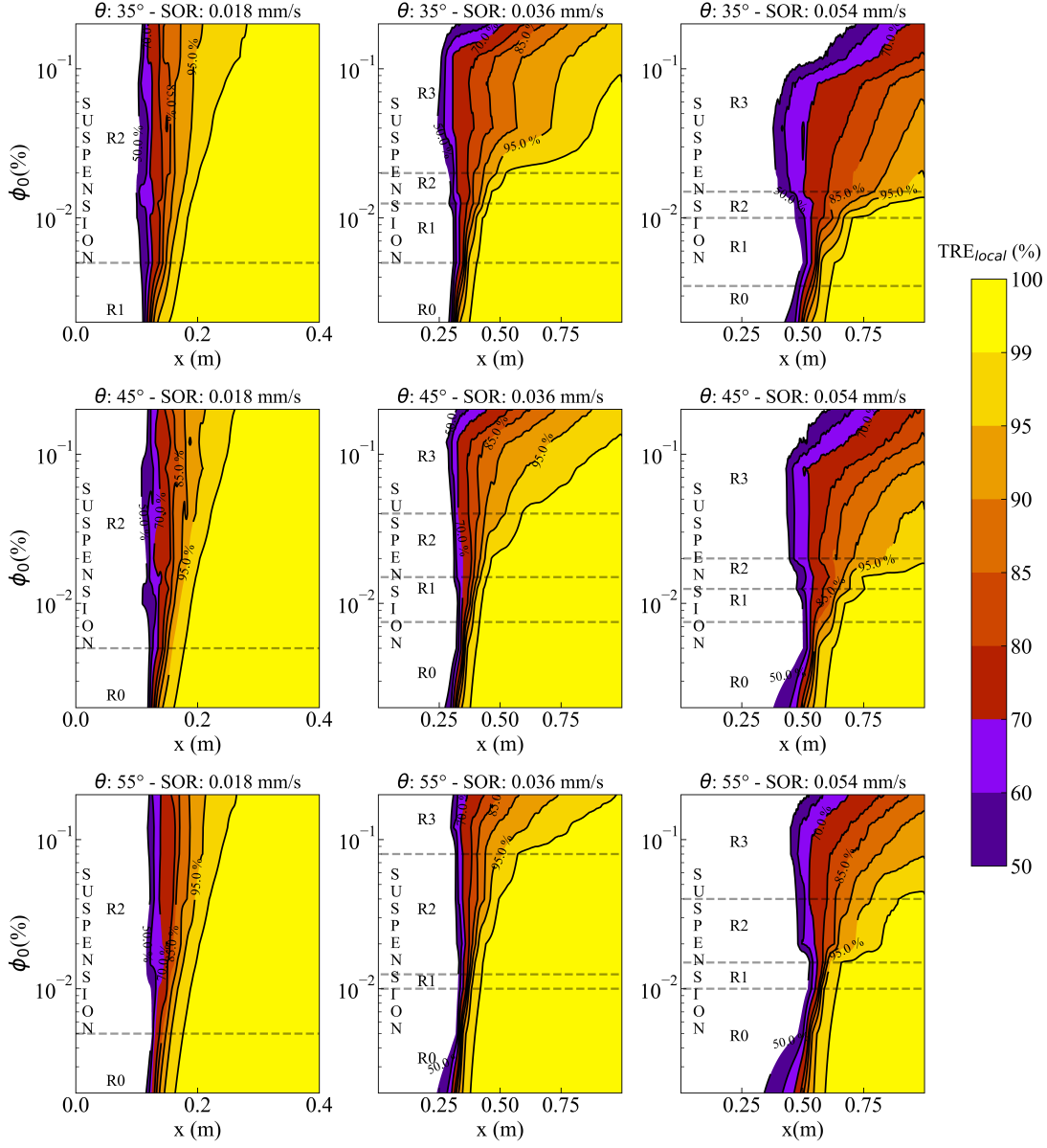}
\caption{Summary panel of the effect of particle concentration on local efficiency (TRE) for different SORs and angles.}\label{f:summary}
\end{figure*}

We have studied the effect of particle concentration systematically for the three SOR and inclination angles described. Figure~\ref{f:summary} presents a summary of local efficiency given its particle concentration for different angles and SOR where the colors refer to the degree of efficiency, with yellow being close to $100\%$  efficiency and purple being poor efficiency. This figure also shows the different regimes (R0 to R3) associated with the instabilities for each of the different cases. 

Within the range studied, this system exhibits two extreme states and two intermediate ones. The first extreme is the dissolved limit (R0), associated with the PNK theory, where sedimentation (Boycott effect) prevails and the dissolved limit model of~\citet{Yao70JotWPCF} can be applied. At the other extreme lies a high-vorticity regime (R3), controlled by high concentration, where resuspension dynamics dominate over sedimentation, and particles report to the settler overflow.  Between these two extremes, there are two intermediate regimes in which the timescales of the sedimentation and resuspension processes are similar. In these regimes, resuspension acts as a particle source from the suspension zone, coexisting with a sedimentation process occurring under a gravitational field of magnitude $g\cos\theta$.

\section{Discussion}\label{sec:discussion}

The results and analysis presented above demonstrate a crucial role on instabilities and resuspension of particle concentration, even while remaining confined to low volume fractions of less than half a percent. 
Previously, \citet{borhan1988sedimentation} studied the effect of particle concentration on the asymptotic solutions first derived in \citet{shaqfeh1986effects} for different levels of inertia (represented by $\mathcal{R}$ mentioned after Eq. \eqref{eq:lambda}) and their stability. 
They found that increasing particle concentration has a stabilizing effect on the clear-fluid\textendash suspension interface, but that increasing inertia destabilizes. 
In his subsequent study, \citet{borhan1989Experiments} conducted experiments with concentrations of the order of $10 \%$ and fluids with viscosities higher than those of water, showing that as the concentration increases, the interface tends to stabilize.  This is in general agreement with the previous results of \citet{borhan1988sedimentation} and seemingly in contradiction with our results. 
However, a different trend is observed in the concentration range between 0 and 0.05 (Figures 8 and 31 in \citet{borhan1988sedimentation}), which is the focus of this study. 
For those low concentrations, destabilization (or no noticeable effect, depending on the viscosity of the fluid) is found for increasing concentration. 
The subsequent stabilization of the system suggests the existence of a critical concentration (higher than those evaluated here) at which the unstable trend leading to resuspension ceases, giving way to stabilization likely driven by the emergence of effective viscosity associated with the increased particle concentration.

Previous studies on the instabilities of these systems \citep{Leung83b_IECPDaD, Probstein77proc, davis1983wave, herbolzheimer1983stability, borhan1988sedimentation, borhan1989Experiments, Brown86thesis} have primarily focused on the development of shear instabilities at the interface between the clear water layer near the upper wall and the suspension. As demonstrated in our results, this type of instability plays a significant role in advancing the resuspension zone toward the overflow. However, it does not account for the destabilizing mechanism observed at the tip of the suspension. At this position, unstable overhanging zones form and give rise to recirculation zones. This mechanism is particularly important, as it directly affects the degree of mixing between the bulk suspension and the clear water. Although it is not the main mechanism associated with the projection of the resuspension zone into the overflow, it is responsible for the degree of mixing in the bulk suspension zone and the concentration profile within it. 
A similar instability mechanism, analogous to Rayleigh-Taylor instabilities, was identified by \citet{zhang2021fluid} for larger particles (\SI{0.6}{mm}) using CFD-DEM simulations. They reported that these instabilities, along with Kelvin-Helmholtz-like instabilities, influence particle settling velocity and exhibit inclination angle-dependent effects.

Identifying the onset of these instabilities, as well as the associated behavioral changes, makes it possible to optimize the selection of the operating concentration or to foresee the impact of an increase in the inlet concentration in settlers of this type. 
As a specific example, the level of fluctuations at the tip of the suspension could be monitored to actively control the feed rate.

\subsection{Lessons learned for the design and operation of lamella settlers}

The results and analysis presented above could contribute to the design of lamellar systems. These findings illustrate how unstable conditions develop as the system processes varying inlet concentrations, significantly impacting its efficiency. According to our findings, the severity of these instabilities correlates with particle concentration. These flow instabilities, in turn, influence the flow profiles within the settler and could have important implications for the design of inclined systems.

\begin{figure*}[!ht]
\centering
\includegraphics[width=0.9\linewidth]{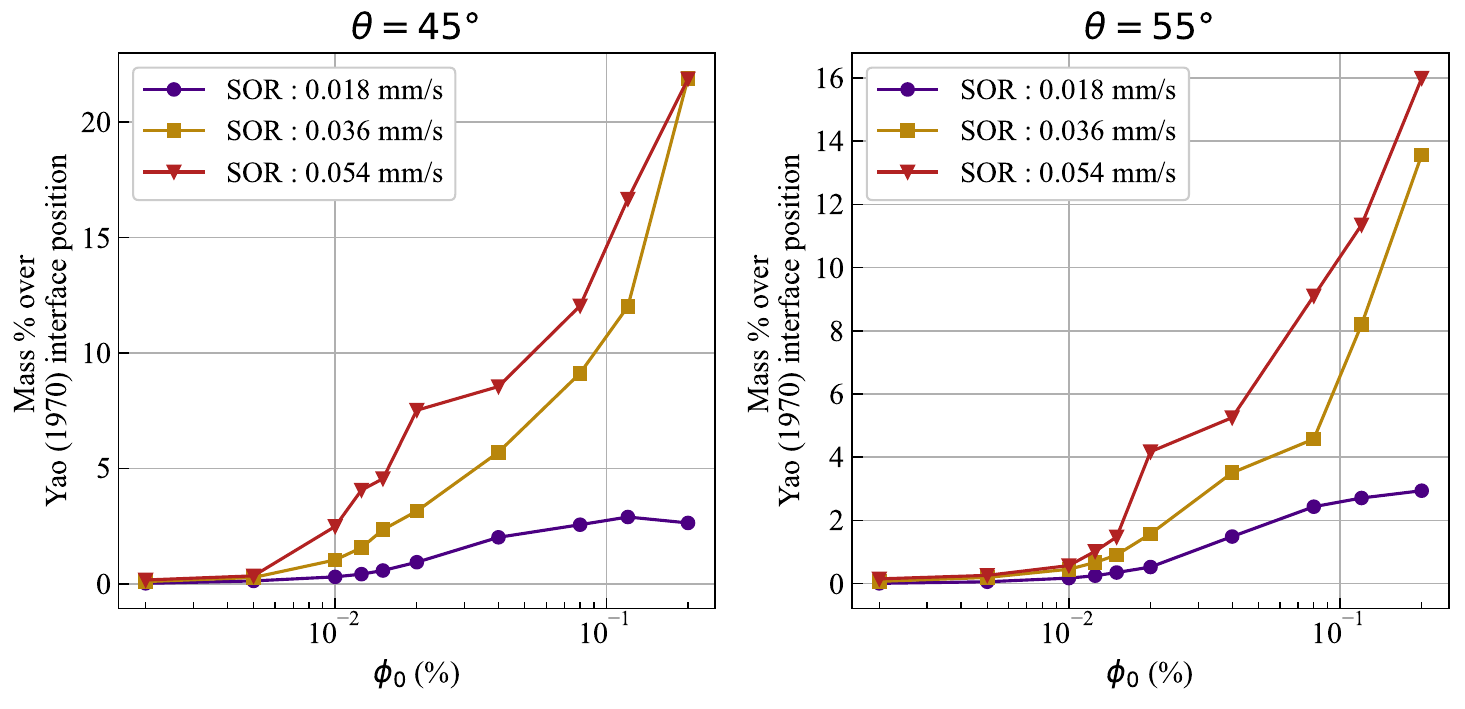}
\caption{Mass percentage over the theoretical line ($L_m$), as mentioned in the Section~\ref{sec:massOverflow}, as a function of inlet concentration $\phi_0$, for $\theta = \SI{45}{\degree}$ (left) and  $\theta = \SI{55}{\degree}$ (right) }\label{f:theorical}
\end{figure*}

Part of the effect on design can be observed by calculating the mass over the theoretical line, as mentioned in section~\ref{sec:massOverflow}. Figure~\ref{f:theorical}, which illustrates the mass fraction above the theoretical line ($L_m$) for two angles, shows that a significant fraction of particles lies above this line, and this fraction increases notably with higher SOR values and a greater fraction of inlet particles. This would indicate that the concentration of small particles ($\approx \SI{10}{ \micro m}$) should be taken into account. An increase in the number of these particles at the inlet can bring with it a considerable loss of efficiency under the same operating conditions. 
Figure~\ref{f:theorical} also highlights the influence of SOR. As shown in Figures~\ref{f:sor_prof} and~\ref{f:summary}, lower SOR values limit instability development compared to higher SORs. Indeed, Figure~\ref{f:theorical} shows that the resuspended mass near the resuspension zone begins to converge for lower SORs, suggesting that the instability does not evolve into a different regime. This likely occurs because a lower SOR reduces the suspension's extent, limiting the available length for wave propagation and amplification. Therefore, SOR, which defines $L_m$, is a key parameter in the evolution of instabilities in these inclined systems.

%% AQUI VOY

For design purposes, low concentrations allow the use of theoretical height assumptions, with models such as the Yao model \citep{Yao70JotWPCF} and PNK theory \citep{ponder1925sedimentation, nakamura1937cause, kinosita1949sedimentation} performing well within this range. However, high concentrations require adjustments by incorporating a resuspension zone into the design calculations. This adjustment depends on the characteristics of the suspension being treated and parameters such as the SOR and the inclination angle, which also influence efficiency. Optimizing the SOR involves balancing throughput and efficiency, while avoiding excessive turbulence that could disrupt the particle separation process. Regarding inclination, steeper angles reduce projected area, requiring longer particle travel paths, increasing the risk of overflow, but enhancing self-cleaning by preventing particle stagnation. Optimizing settler performance involves evaluating concentration, flow parameters, and geometry to enhance efficiency and identify limitations based on suspension properties.

\subsection{Instability route}

As mentioned previously, the existing literature has focused predominantly on shear instability within inclined systems. However, according to the results presented in this study, we observe that the instability begins at the tip of the suspension region. 
As described in \S \ref{sec:resultsConcentration}, the destabilization process is associated to different regimes of flow separation, which is a common fluid mechanical phenomenon. 
Indeed, flow separation occurs at a sudden widening of a channel depending on a suitably defined Reynolds number; the clear-fluid layer has a sudden expansion at the tip of the suspension region. Furthermore, a parallel between the different regimes can be drawn to a prototypical case of flow separation, that of the flow around a cylinder \citep{Frisch1995}. 
For very low Reynolds numbers, the flow remains attached from the front to the back of the cylinder. 
At a slightly larger Reynolds number ($\approx 10$), the flow separates and a recirculation region appears and remains attached to the cylinder, similar to the recirculation region that appears in the wall and remains attached to the suspension in Regime 1. 
Further increasing the Reynolds number beyond $\approx 47$, the Bénard\textendash von-Kármán vortex shedding instability takes place and gives place to a regular pattern of alternating vortices. 
In the present case, baroclinic effects and buoyancy prevent the corresponding vorticity release, but a parallel can be drawn with the pattern of release of recirculation regions from the suspension of Regime~2. An increased Reynolds number leads to further destabilization, turbulence, and mixing, analogous to Regime~3.

These observations suggest that the destabilization process could be characterized by a Reynolds number based on the velocity and widening of the clear-fluid region, that is,
\begin{equation}
    Re_{\bar{\delta}_{susp}} = \frac{\bar{\delta}_{susp}\textbf{U}_{max}}{\nu} = \frac{(B-\bar{\delta})\cos\theta\textbf{U}_{max}}{\nu}
\end{equation}
where $\bar{\delta}_{susp}$ is the horizontal projection of the width of the suspension layer, $\nu$ is the kinematic viscosity of water and $\textbf{U}_{max}$ the local maximum flow velocity. Considering that the suspension layer occupies an approximate width of $B-\delta$, its projection onto a horizontal plane (with the vertical axis defined by gravity) is given by $(B-\delta)\cos(\theta)$. 

As presented in the results, the instability of the system is associated with thinner clear water thicknesses or a wider suspension layer. On the other hand, as we will see in the scaling, the maximum velocity is a characteristic velocity that scales well with the parameters of the inclined system.

\begin{figure*}[!ht]
\centering

    \includegraphics[width=0.95\linewidth]{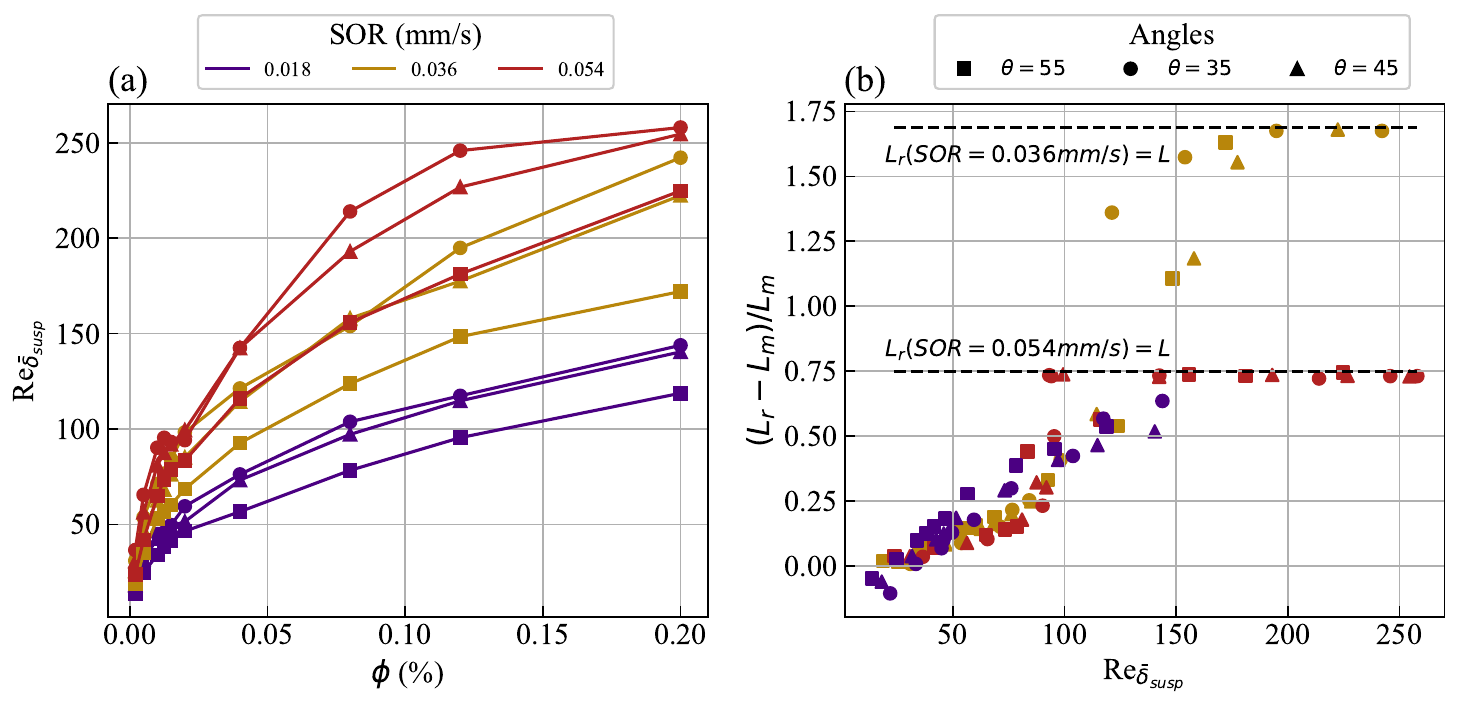}

\caption{(a) $Re_{\bar{\delta}_{susp}}$ for different SOR and inclination angle as a function of particle concentration and (b) The normalized difference between the position of the resuspension zone ($L_r$) and the theoretical position ($L_m$) is presented as a function of $Re_{\bar{\delta}_{susp}}$, for the three angles and SOR. The horizontal lines correspond to the position where resuspension reaches the cell length \(L\) for a given SOR.}
     \label{fig:Re_delta}
\end{figure*}

 Figure~\ref{fig:Re_delta}(a) illustrates how the value of this number increases with higher inlet concentrations, indicating a more unstable system. Furthermore, the relationship between the SOR and the angle of inclination with $Re_{\bar{\delta}_{susp}} $ is evident. Systems characterized by higher SOR values and an angle of inclination of $\theta = \SI{35}{\degree}$, which exhibit greater instabilities and a higher degree of resuspension, also show higher values of this number. In Figure~\ref{fig:Re_delta}(b), the normalized difference between the resuspension zone position ($L_r$) and the theoretical position ($L_m$), i.e. $\frac{L_r - L_m}{L_m}$,
is plotted as a function of $Re_{\bar{\delta}_{susp}}$. The results indicate that while an increase in $Re_{\bar{\delta}_{susp}}$ generally corresponds to a higher degree of resuspension, the effect varies with the SOR. In the case of a higher SOR, \(L_r\) quickly reaches the overflow zone (\(L_r = L\)), whereas for a lower SOR, due to the system being oversized for this SOR, resuspension does not reach the end of the cell.

%\begin{figure}[!ht]
%     \centering
%     \includegraphics[width=1.0\linewidth]{RedeltaS.eps}
%     \caption{$Re_{\bar{\delta}_{susp}}$ for different SOR and inclination angle as a function of particle concentration.}
%     \label{fig:Re_delta}
% \end{figure}

Such an approach allows for a more detailed analysis of the flow and the transition to complex regimes, bringing new perspectives to the understanding of these systems.

\subsection{Particle diameter effect}\label{sec:diameterEffect}

\begin{figure*}[!ht]
\centering
    \begin{tabular}{c}
         \includegraphics[width=16cm]{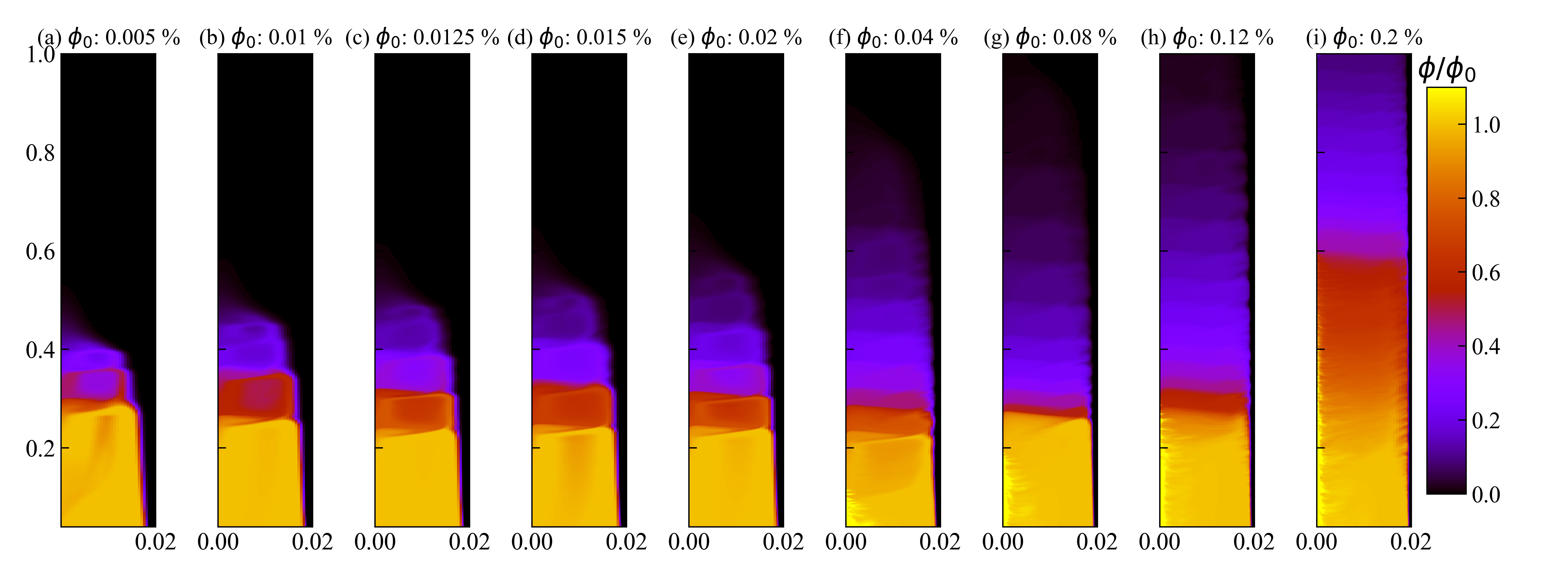} \\
         \includegraphics[width=16cm]{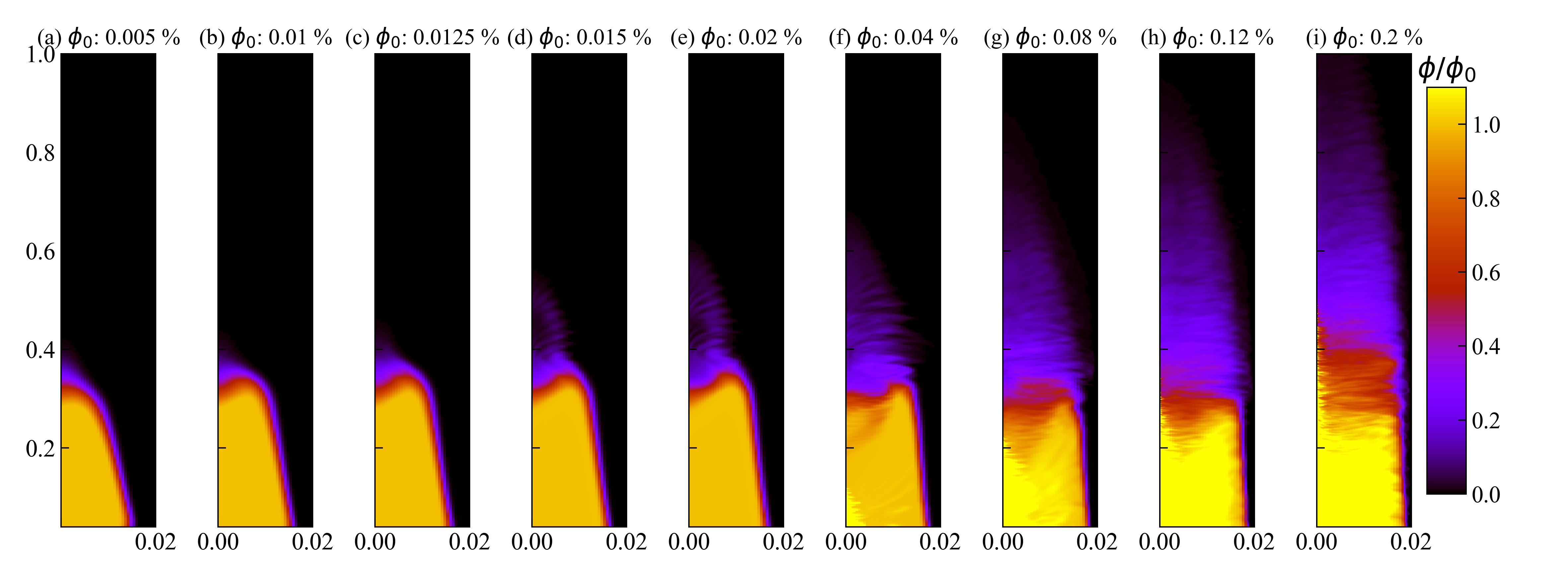}
    \end{tabular}
\caption{Particle concentration profile for different inlet concentrations, $\theta = \SI{45}{\degree}$. Top row corresponds to $d = \SI{5}{\mu m}$, SOR =$\SI{0.009}{mm/s}$  and bottom row to $d = \SI{20}{\mu m}$, SOR =$\SI{0.144}{mm/s}$.}\label{f:diameterEffect}
\end{figure*}

Since in reality suspensions are not monodisperse and present a size distribution, it is relevant to study the behavior at different particle sizes. It is necessary to understand whether instabilities occur in the same way at all sizes or whether the system behaves completely differently. Due to the complexity of analyzing polydisperse cases, it was studied how the present system would be when the diameter is half and double its base size, 5 and 20 microns, respectively.
With respect to the effect of particle size on the behavior of instabilities caused by particle concentration, it can be observed that increasing particle size delays the appearance of the instability that causes plumes near the suspension interface. In the opposite case, at a smaller particle size, instability is present from the first case studied with slow recirculation similar to Regime~1. Both cases can be observed in Figure~\ref{f:diameterEffect}, where the 2D concentration profile can be seen for both diameters.  This is consistent with the fact that a larger particle size is associated with a greater thickness of the clear water layer, allowing the clear water to flow in a less confined way, thus reducing the erosion of the suspension surface. On the other hand, a smaller diameter leads to a smaller thickness and the development of instability for a lower concentration. 
Furthermore, a higher sedimentation rate is not related to an earlier onset of instability. However, the development of instabilities is qualitatively distinct and strongly dependent on this rate. At lower sedimentation rates, the flow is dominated by small-scale recirculation and the formation of a staircase structure, while at higher sedimentation rates, erosion becomes the dominant mechanism. As in the case of the 10~microns, in these cases it is observed that there is a relationship between the thickness of clear water, resuspension and the presence of instabilities in the system. In addition, it is observed that very small particles would generate an increase in resuspension and a greater loss of efficiency in these inclined systems. This is also mentioned by \citet{zhang2021fluid}, who state that particle resuspension is more likely to occur with smaller particle sizes. Regardless of these differences, it is confirmed that in all cases the destabilization begins at the tip of the suspension as the concentration increases.

Other studies report similar observations. \citet{chang2019three} show that instability depends on both particle size and inclination angle. Larger particles can hasten turbulence onset, although overall stability varies with both factors. For example, larger diameters (\SI{150}{\micro m}) can be more stable than smaller ones (\SI{100}{\micro m}). They also note that shear instability arises from the competition between buoyancy-induced flow and clear water layer thickness.  Similarly, \citet{chang2021transient} found that increased settling velocity and horizontal inclination enhance stability.  These findings underscore the critical role of particle size in system stability.

\subsection{Scaling of the thickness and velocity of the clear-fluid layer}

\begin{figure*}[!ht]
    \centering
    % Primera fila con una subfigura
    \begin{subfigure}[b]{0.55\textwidth}
        \centering
        \includegraphics[width=\textwidth]{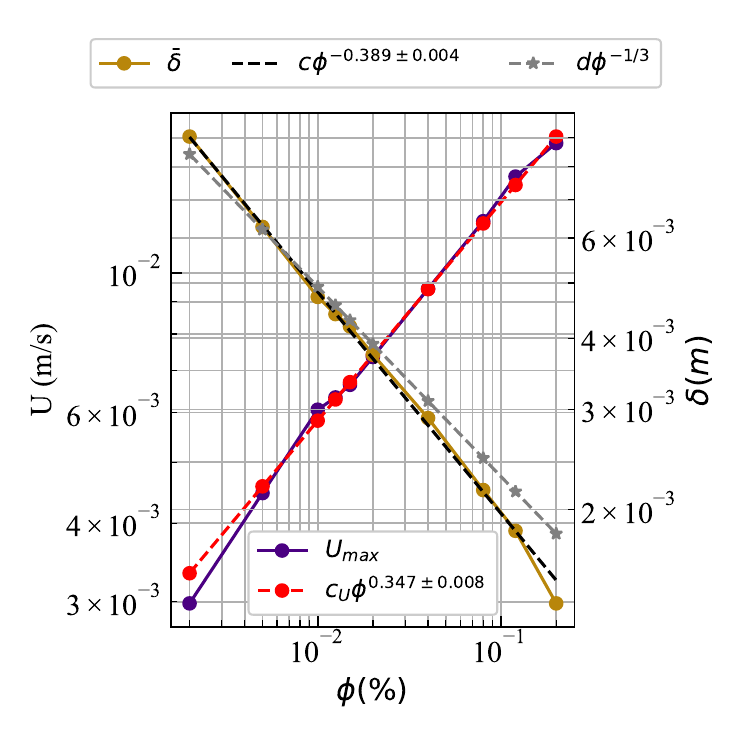}
        \caption{SOR = 0.036 mm/s}
        \label{fig:subfigura1}
    \end{subfigure}

    % Segunda fila con dos subfiguras
    \begin{subfigure}[b]{0.48\textwidth}
        \centering
        \includegraphics[width=\textwidth]{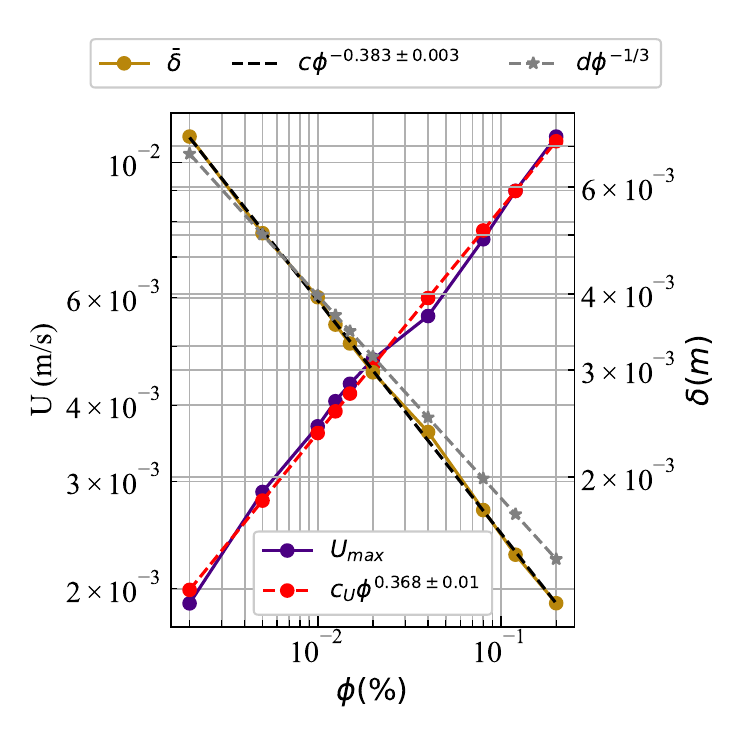}
        \caption{SOR = 0.018 mm/s}
        \label{fig:subfigura2}
    \end{subfigure}
    \hspace{0.02\textwidth}
    \begin{subfigure}[b]{0.48\textwidth}
        \centering
        \includegraphics[width=\textwidth]{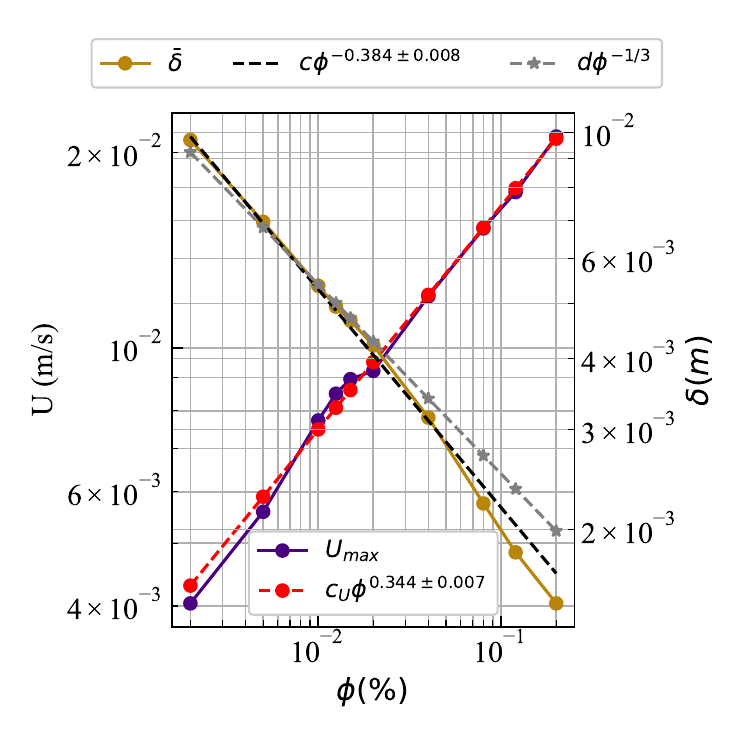}
        \caption{SOR = 0.054 mm/s}
        \label{fig:subfigura3}
    \end{subfigure}

    \caption{Velocity and layer thickness of clear water for three different SOR, $\theta = \SI{55}{\degree}$, as a function of particle concentration. The gray dashed curve represents the fitted curve ($c\phi^{b}$), while the black dashed curve corresponds to the curve $d\phi^{-1/3}$ (value after $\pm$ is the parameter fit error). The value for $c$ and $d$ for different SOR (with its respectively curve Root Mean Squared Error) are: (a) $c = 1.33\times10^{-4}$ (RMSE = $6.8\times10^{-5}$) and $d = 2.2\times10^{-4}$ (RMSE = $3\times10^{-4}$), (b) $c = 1.54\times10^{-4}$ (RMSE = $3.7\times10^{-5}$)  and $d = 2.5\times10^{-4} $ (RMSE = $2.1\times10^{-4}$) and (c) $c = 1.14\times10^{-4}$ (RMSE = $1.3\times10^{-4}$) and $d = 1.8\times10^{-4}$ (RMSE = $3\times10^{-4}$).  }
    \label{fig:Umax_espesor}
\end{figure*}

The clear relationship between particle concentration and efficiency loss has several factors influencing this behavior, which include the clear water thickness and the maximum velocity. Simulations show that the thickness of clear water decreases as the particle concentration increases, directly reducing the space available for the water to exit the suspension zone. This reduction is accompanied by an increase in the maximum velocity with higher concentrations.

Log-log plots of the maximum velocity and thickness of the clear water layer versus particle concentration for three different surface overflow rates are shown in Figure~\ref{fig:Umax_espesor}, accompanied by a fitted model of the form ($C\phi^{b}$) to determine the exponents relating these parameters. The results indicate that the exponent for the thickness fit ranges from $-0.37$ to $-0.4$, while for the velocity, it varies from $0.34$ to $0.37$. Additionally, the figure reveals that maximum velocity can increase up to sixfold and thickness up to tenfold depending on the particle concentration.

%Both exponents are close to the theoretical value of 1/3 obtained from scaling, especially in the case of velocity.

\begin{figure*}[!ht]
     \centering
     \includegraphics[width=1\linewidth]{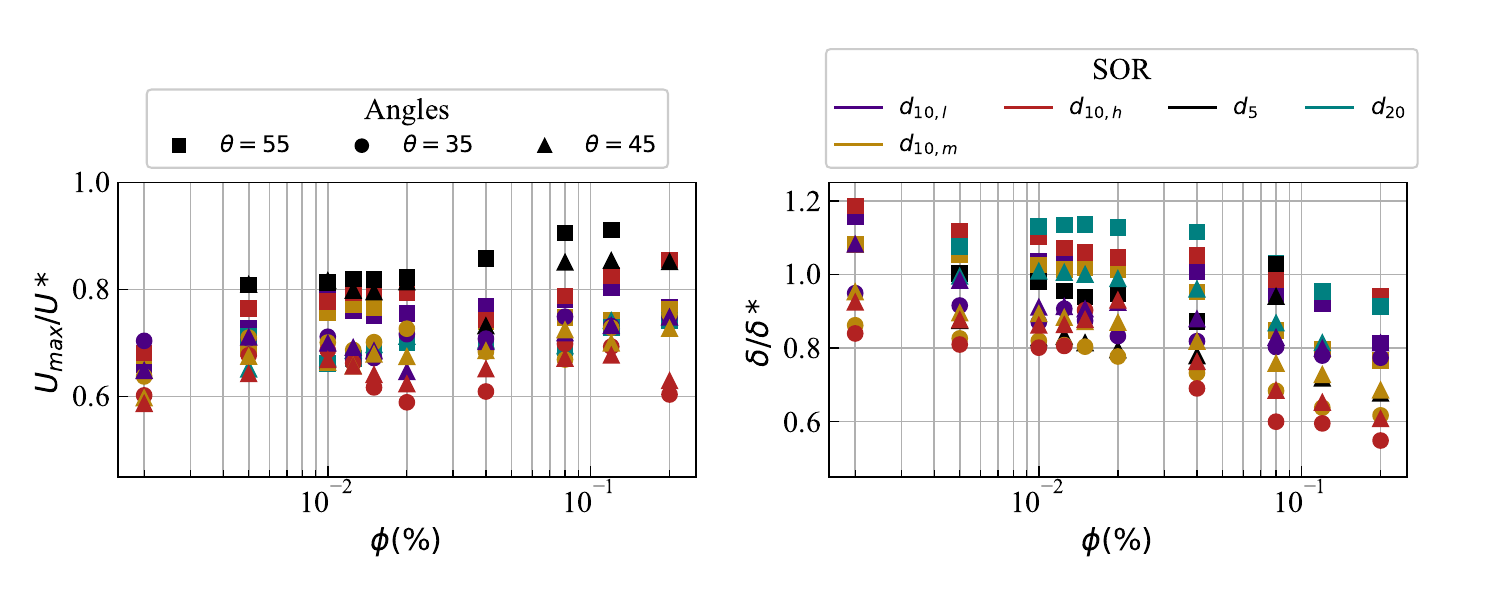}
     \caption{Scaled values of thickness and maximum velocity, $\theta = 35,45,\SI{55}{\degree}$. For three different SOR. The subscript l, m and h means low, medium and high SOR respectively (see Table~\ref{t:mainParams}). SOR for $d = \SI{5}{\micro m}$ and $d = \SI{20}{\micro m}$ was 0.009 and 0.144 mm/s respectively. }
     \label{fig:scalingUdelta}
 \end{figure*}

 According to the theoretical scaling pointed out by \citet[\S 5.2.1]{Blanchette03thesis} (see also \citet{reyes2022review}), the scales of the longitudinal velocity and the clear water thickness correspond to:
\begin{align}
    \delta^* \sim \mathcal{L}\Lambda^{-1/3}, \label{eq:scalingdelta} \\
    U^* \sim W_s\Lambda^{1/3},  \label{eq:scalingU}
\end{align}
with $\Lambda$ calculated by equation ~\eqref{eq:lambda}. In the continuous case, the length scale ($\mathcal{L}$) is defined by the steady-state suspension position, which, in our study, corresponds to $L_m$ as calculated by equation~\eqref{eq:LmDef} . Therefore, the thickness and velocity scales can be expressed as:
 \begin{align}
     \delta^* \sim  \left(L_m a^2\right)^{1/3}\phi^{-1/3}, \label{eq:delta*}\\
     U^* \sim \left(\frac{W_s^3 L_m^2}{a^2}\right)^{1/3} \phi^{1/3}.  \label{eq:U*}
 \end{align}
On these scales the maximum velocity and the average thickness yields the values presented in Figure~\ref{fig:scalingUdelta}. Here, the subscripts l, m and h denote low, medium, and high SOR values, corresponding to \SI{0.018}{mm/s}, \SI{0.036}{mm/s}, and \SI{0.054}{mm/s}, respectively. The results show that the scaled velocities and thicknesses for different angles, SOR, and particle diameter converge within a relatively narrow range of values, showing that the scaling effectively captures the trends of these parameters. 

However, for the thickness of the clear water, a correction in the scaling appears necessary, since the fit begins to diverge from the model $D\phi^{1/3}$ (see gray line in Figure~\ref{fig:Umax_espesor}) at higher concentrations, indicating that this scaling may not be accurate at elevated concentrations (see also Figure~\ref{fig:Umax_espesor}).
An option to adjust the velocity scale could be the use of a hindered settling velocity relationship, such as the Richardson-Zaki model \citep{richardson1997sedimentation}. 
Within the studied range, well below $\SI{0.5}{\%}$, this effect is negligible, but should be considered at higher concentrations. 
A more likely explanation is that the scalings in Eqs. (\ref{eq:scalingdelta}-\ref{eq:scalingU}) are valid for self-similar solutions with fixed $\mathcal{R}^{1/2}$, indicating a fixed level of inertia as $\Lambda\rightarrow\infty$ \citep{shaqfeh1986effects}. 
However, as concentration increases, the flow appears to approach more viscous solutions, \emph{i.e.}, with lower $\mathcal{R}$ corresponding to decreasing levels of inertia.
This is consistent with the displacement of the position of the maximum velocity with respect to the clear-fluid\textendash suspension interface (insets in Figure~\ref{f:stream}), which can be compared with results in \citet{shaqfeh1986effects, borhan1988sedimentation} showing the change of the velocity profile depending on inertia. However, the scales \eqref{eq:delta*} and \eqref{eq:U*} allow us to capture the dominant behavior of velocity and thickness in this system for this concentration range.

\section{Study Limitations}

This study has certain limitations that should be acknowledged. First, the particle size range was restricted to \num{5}--\SI{20}{\micro m}. Larger particles, which tend to induce more turbulent flow, would necessitate a three-dimensional analysis. Additionally, wall roughness and interparticle cohesive forces were not considered, which could affect sediment bed development.  Moreover, while the sediment bed was generally thin ($< \SI{1}{mm}$), higher inlet concentrations can lead to rapid sediment bed growth and increased thickness.  A complementary study focusing on the sediment layer and its interaction with the suspended and clear water layers would enhance understanding of these systems.  Finally, the concentration range examined (\num{0.002}--\SI{0.12}{\%}) could be extended to higher values to investigate the impact of mixture viscosity on instability development.

\section{Conclusions}

The present study underscores the critical role of particle concentration, particularly for fine particles, in the design and operation of lamellar settling systems. Within the range studied, different regimes (R0-R3) were identified, each associated with different behavior patterns of the system. Starting with a stable system (R0), as concentration increases, the system transitions to developing instabilities in the suspension region, characterized by unstable overhanging zones that lead to recirculation (R1). With further increases in concentration, multiple recirculation zones emerge, forming a staircase structure (R2). At higher concentrations, the resuspension zone tends to expand, resulting in a significant degree of resuspension driven by the increasing relevance of shear-induced instabilities (R3). The concentration at which each of these mechanisms occurs depends on other parameters, such as the SOR and the angle of inclination. Elevated SOR values induce turbulence, accelerating instability effects, while steeper inclinations (more vertical cells) reduce the projected sedimentation area, promoting the onset of instabilities at lower concentrations. Understanding the effect of each parameter on resuspension is crucial, as it directly influences the solid-liquid separation efficiency in these inclined systems.

Although the effect of particle size was not explored in depth, it was shown that particle size influences the behavior of instabilities within the same concentration range. For smaller particle sizes, recirculation zones dominate, whereas for larger particle sizes, erosion becomes the prevailing mechanism beyond the concentration at which the system becomes unstable.

This study also introduced a Reynolds number based on the thickness and maximum velocity, along with velocity and thickness scales derived from the theoretical length, incorporating the effects of SOR, inclination angle, and particle size. It is also a dimensionless number that depends on internal characteristics of the flow, and would require either numerical simulation or experimental work to be identified. In parallel, the velocity and thickness scales presented in the discussion provide a practical framework for approximating the behavior of the system. However, adjustments are required to consider the effect of particle concentration. These findings could serve as a reference for future analyses of similar systems.

Although this study provides valuable information on the influence of particle concentration on solid-liquid separation in inclined settlers, several knowledge gaps remain. In particular, further research on the effect of the mean particle diameter would improve our understanding of the behavior of the present system. Furthermore, the present study employs a simplified 2D model, which may not fully capture the complex 3D flow structures and particle interactions that occur in actual settlers, and particularly regarding the effect of particle size and concentration on lateral particle distributions. Future research could explore 3D simulations to provide a more comprehensive understanding of the dynamics of the system. Finally, experimental validation of numerical results, particularly in relation to the onset and development of instabilities, would strengthen the conclusions drawn from this study and provide validation of the current results for the development of future engineering concentration-dependent models.

\begin{acknowledgments}
The authors gratefully acknowledge support from the Department of Mining Engineering of the University of Chile, the Chilean National Agency for Research and Development (ANID) through projects Fondecyt 1211044, Anillo ACT210027, Fondef ID23I10333, and National Doctoral Scholarship 21200441. 
CA acknowledges support of Nordita and the Swedish Research Council Grant No. 2018-04290.
\end{acknowledgments}

%\nocite{*}
%\bibliography{references}% Produces the bibliography via BibTeX.
%

%%%%%%%%%%%%%%%%%%%%%%%%%%%%%%%%%%%%%%%%%
%%%%%%%%%%%%%%%%%%%%%%%%%%%%%%%%%%%%%%%%%
\appendix
%\color{red}

\section{Mesh Resolution}
\label{appendix:mesh}

Here, we present the details of the mesh resolution and convergence studies.

%\subsection{Mesh and Convergence to steady state}
%\subsubsection{Mesh Comparison}
\subsection{Mesh Comparison}

\begin{table}[!ht]
\centering
\caption{Number of cells of different mesh sections. Sections are shown in Figure \ref{f:system}.}
\label{t:mesh_val}
\begin{ruledtabular}
    \begin{tabular}{ p{1.1cm}p{1cm}p{1.6cm}p{1.3cm}p{1.6cm}p{1.3cm}  }
     Mesh  & Center &Longitudinal & Boundary layer (Left) & 
 Boundary layer (Right) & CPU time (hrs)\\
     \hline
     Coarsest & 18 & 880 & 30 &10 & 30.4\\
    Coarse & 24 & 1183 & 30 & 12 & 30.3\\
    Medium & 30 & 1583 & 30 & 15 & 92.8 \\
    Fine & 40 & 2000 &  30 & 20 & 133.8\\
    Finest & 50 & 3000 & 30 & 30 & 202.6\\
    \end{tabular}
\end{ruledtabular}
\end{table}

The mesh used in the simulations was a structured 2D grid with quadrilateral elements, which has been shown to have better convergence and higher resolution than an unstructured grid \citep{chawner2013quality}. The formation of the suspension, clear fluid and sediment layer (the last two of them very thin), requires to shape a fine mesh close to the wall. This refinement is crucial for the appropriate development of boundary layers, which play a significant role in this system as they govern mixing, resuspension, and the gravity flow towards the hopper.
To study the dependence of the density of the mesh, five meshes were considered: coarsest, coarse, medium, fine, and finest (see Table \ref{t:mesh_val}), which have different levels of refinement in the cells of the boundary layer and in the center cells. In all cases, the boundary layer close to the bottom boundary of the cell was highly refined so that the sediment layer (the thickness of which is less than 1 mm) is correctly developed.
It is important to note that the aspect ratio considered in the center cells was close to 1 to avoid a possible gradient preference. Generally, aspect ratios greater than 1 are used near the boundary layer, where normal gradients are known to be particularly higher than those tangential to the wall \citep{greenshields2022notes}. On the other hand, in areas of high vorticity, where the velocity vector changes constantly and also the velocity gradient varies, cell aspect ratios close to 1 have been chosen to avoid abrupt variations of $\Delta t$, in order to maintain the Courant number or loss of accuracy.

\begin{figure}[!ht]
\centering

         \includegraphics[width=8.cm,keepaspectratio]{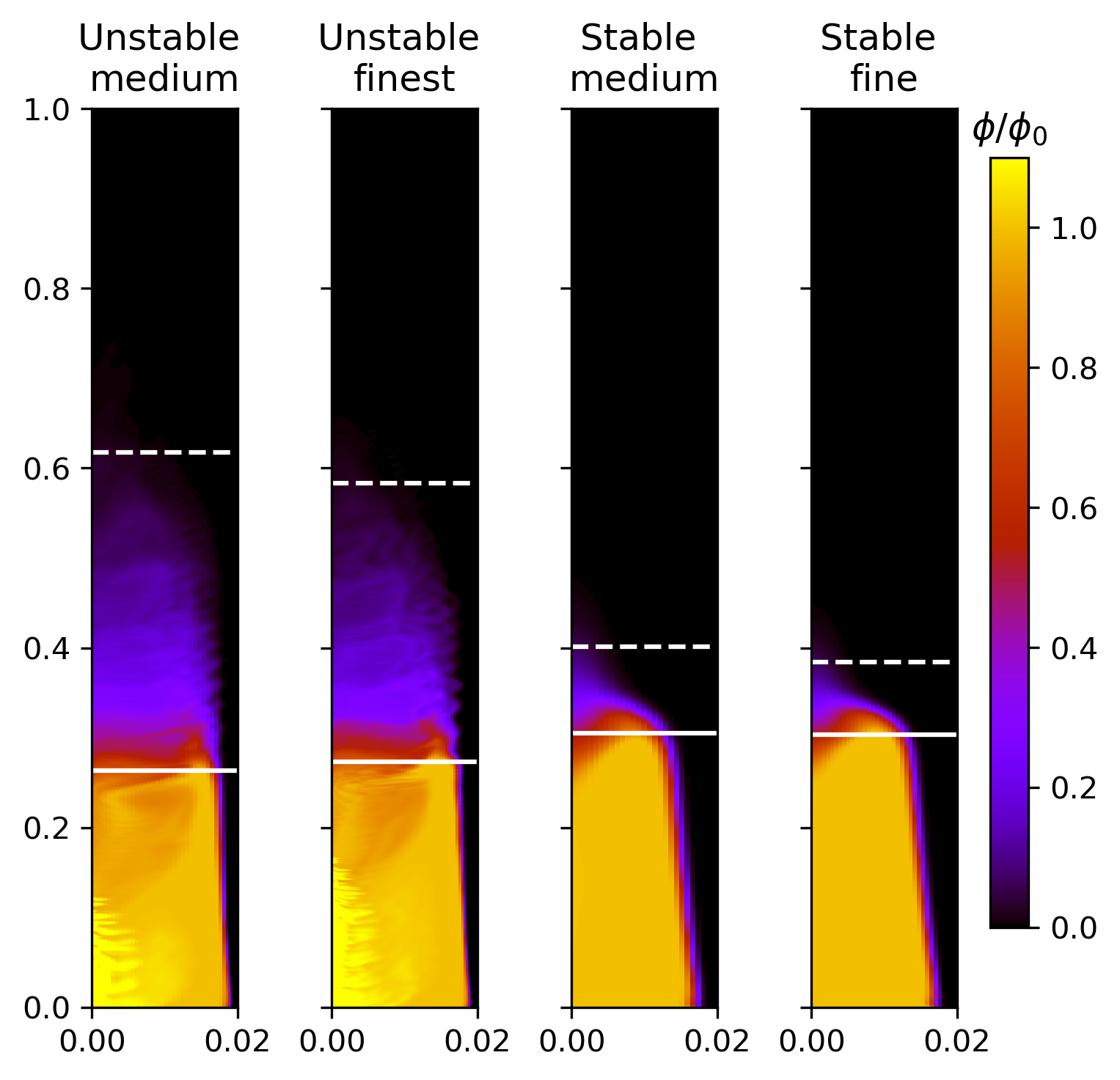}
    
\caption{2D particle concentration profile for stable ($\phi_0=0.005\%$) and unstable ($\phi_0=0.05\%$) cases, where two levels of refinement are shown (left). Solid white line corresponds to suspension bulk position ($L_b$) and dashed line to resuspension position ($L_r$). }\label{f:mesh_aspects_l}
\end{figure}

\begin{figure}[!ht]
\centering
         \includegraphics[width=8.cm]{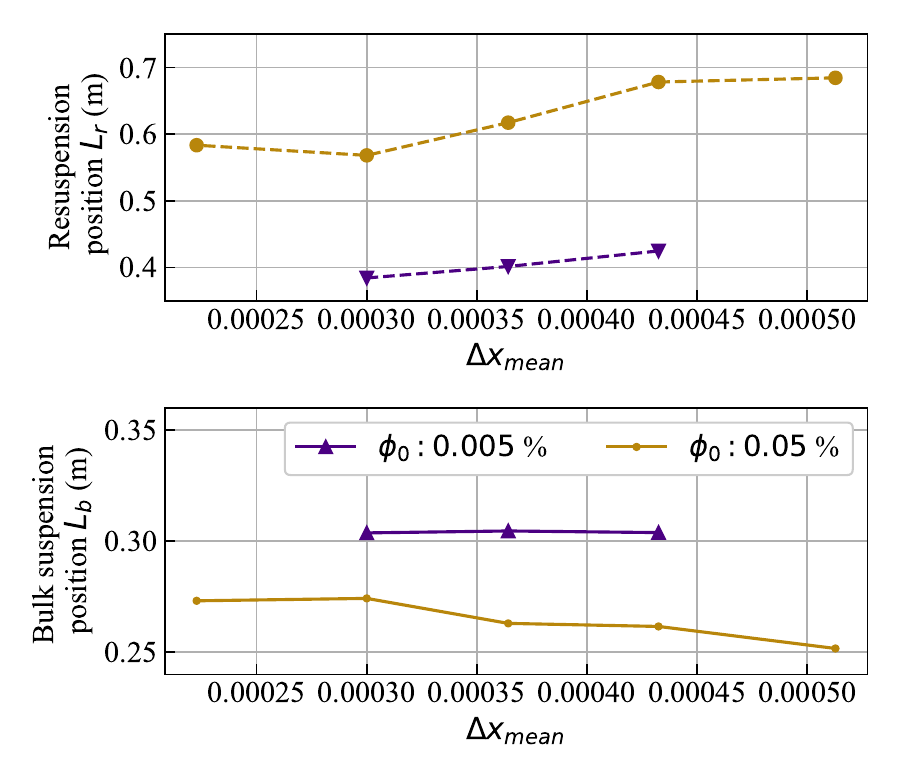}
\caption{ Mesh dependence of the suspension and resuspension positions, where $\delta x_{mean} = (1/n)\sum \sqrt{A_i}$, with $A_i$ representing the area of cell $i$.}\label{f:mesh_aspects_r}
\end{figure}

%\begin{figure*}[!ht]
%\centering
%    \begin{tabular}{cc}
%        (a) & (b) \\
%         \includegraphics[width=8.cm,keepaspectratio]{mesh_comparison.png}&
%         \includegraphics[width=8.cm]{deltax_vs_xV2.eps}
%    \end{tabular}
%\caption{2D particle concentration profile for stable ($\phi_0=0.005\%$) and unstable ($\phi_0=0.05\%$) cases, where two levels of refinement are shown (left). Solid white line corresponds to suspension bulk position ($L_b$) and dashed line to resuspension position ($L_r$). On the right, mesh dependence of the suspension and resuspension positions, where $\delta x_{mean} = (1/n)\sum \sqrt{A_i}$, with $A_i$ representing the area of cell $i$.}\label{f:mesh_aspects}
%\end{figure*}

For this mesh analysis, two cases were studied: with and without the onset of flow instabilities. 
This was necessary because mesh density requirements are expected to vary because the physics in unstable cases differs from that in stable cases. 
Both cases are observed in Figure~\ref{f:mesh_aspects_l}, where it shows the 2D profile of the particle concentration for different levels of refinement, and the right side shows the interface positions for bulk suspension ($L_b$) and resuspension ($L_r$) as a function of the mean size of the cells $\Delta x_{mean}$. 
As can be observed in the variation of the interface positions $L_b$ and $L_r$ (see Figure \ref{f:mesh_aspects_r}), the dependence on the mesh is greater for unstable cases (with physical instabilities), associated to higher particle concentrations. 
Note that the position of the resuspension $L_r$ is more sensitive to the mesh, in part because it strongly depends on the development of instabilities, which are sensitive to the mesh resolution. % and a finer mesh allows the development of smaller eddies.
Although the position of the bulk suspension is mainly determined by the inflow rate and the sedimentation velocity of the particles, it is also influenced by the erosion of the suspension due to instabilities. 
For stable cases, the bulk suspension positions are similar, whereas there is a slight difference for the resuspension interface. 
%For these unstable cases, 
Qualitative similarities can be observed in all cases; however, from the medium mesh onward, similar profiles are observed, both in particle concentration (see Figure~\ref{f:mesh_aspects_l}) and flow velocities.

Overall, according to the results of the different meshes, decreasing differences can be observed as the mesh becomes finer, especially from the medium mesh onward. However, as expected, decreasing cell size has a direct effect on computational time. The computational time is %roughly about four times higher 
almost seven times larger when the average cell size is reduced by a factor of two below the coarse mesh (see Table~\ref{t:mesh_val}, last column). For this reason, the medium mesh density was selected for this study as a compromise between accuracy and the ability to run multiple cases to analyze different aspects of this system. 

%\subsubsection{Initial conditions and convergence to a statistically steady state}\label{sec:convergence}

\section{Solution procedure}\label{ap:procedure}

\begin{figure*}[!ht]
     \centering
     \includegraphics[width=1\linewidth]{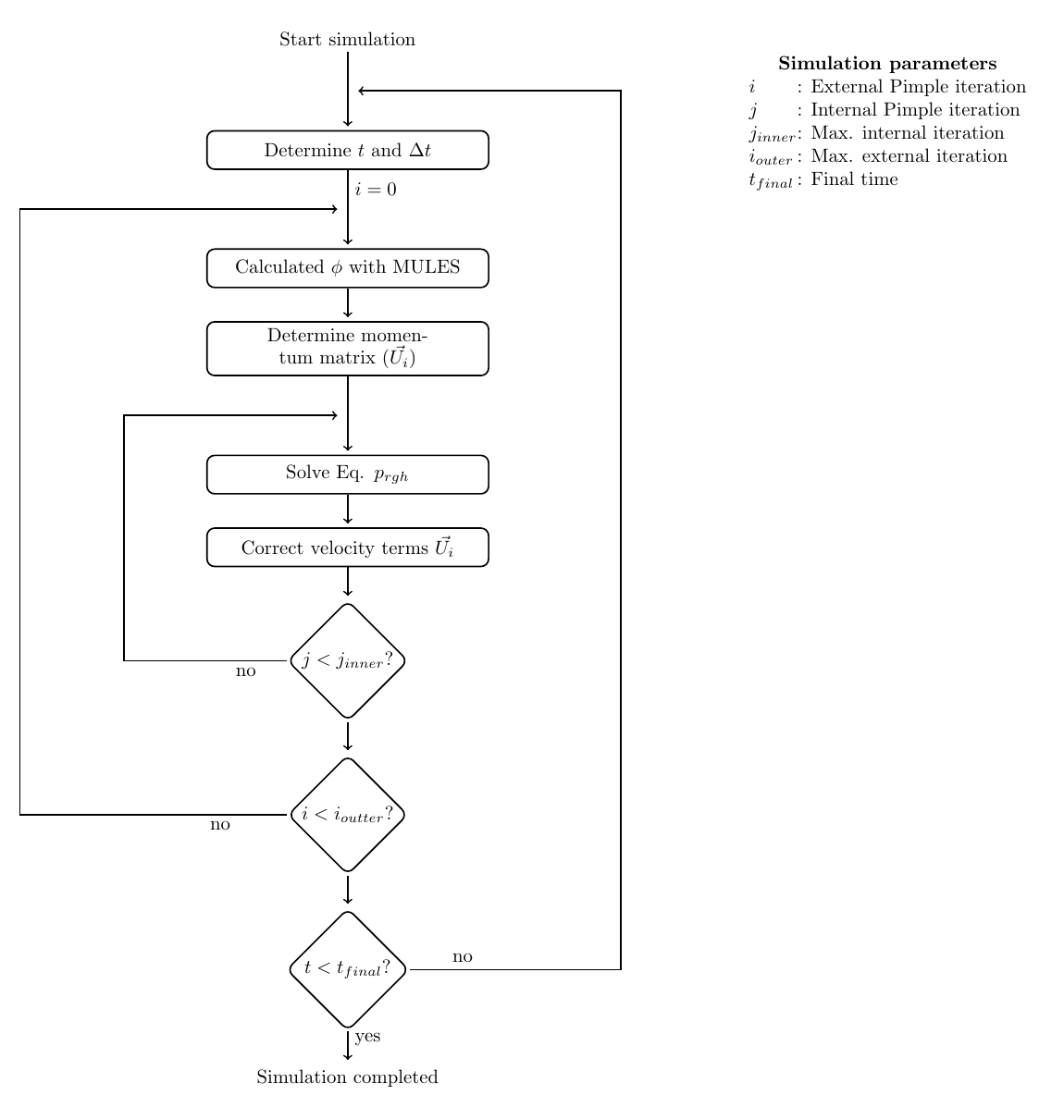}
     \caption{Flowchart illustrating the solution procedure using the PIMPLE algorithm, a hybrid method that integrates the SIMPLE and PISO approaches for robust and accurate transient multiphase flow simulations.}
     \label{fig:diagram}
 \end{figure*}

To provide a clear understanding of the computational workflow, this appendix includes a flowchart detailing the solution procedure used in the CFD simulations. This diagram (Figure~\ref{fig:diagram}) outlines the steps of the numerical approach.

\color{black}

\end{document}